\documentclass[aps,pra,a4paper,reprint,twocolumn,floatfix,superscriptaddress,notitlepage,usenames,dvipsnames,svgnames,table,nofootinbib,longbibliography]{revtex4-1}
\pdfoutput=1
\usepackage[utf8]{inputenc}
\usepackage[T1]{fontenc}
\usepackage{graphicx}
\usepackage{grffile}
\usepackage{wrapfig}
\usepackage{rotating}
\usepackage[normalem]{ulem}
\usepackage{amsmath}
\usepackage{textcomp}
\usepackage{amssymb}
\usepackage{capt-of}
\usepackage{verbatim}

\usepackage{parskip}
\usepackage{mathrsfs}
\usepackage[margin= 0.75in]{geometry}
\usepackage[braket, qm]{qcircuit}

\usepackage[english]{babel}
\usepackage{letltxmacro}
\LetLtxMacro{\ORIGselectlanguage}{\selectlanguage}
\makeatletter
\DeclareRobustCommand{\selectlanguage}[1]{%
  \@ifundefined{alias@\string#1}
    {\ORIGselectlanguage{#1}}
    {\begingroup\edef\x{\endgroup
      \noexpand\ORIGselectlanguage{\@nameuse{alias@#1}}}\x}%
}
\newcommand{\definelanguagealias}[2]{%
  \@namedef{alias@#1}{#2}%
}
\makeatother
\definelanguagealias{en}{english}

\usepackage[autostyle,english=american]{csquotes}
\MakeOuterQuote{"}

\usepackage{bbm}
\usepackage{etoolbox}
\usepackage{tcolorbox} 
\usepackage{fleqn} 
\usepackage{tikz}
\usepackage{amsthm}
\usepackage{amssymb}
\usetikzlibrary{arrows.meta}
\usepackage{mathtools}

\newcommand{\prlsection}[1]{{\em {#1}.---~}}

\usepackage{bm}
\usepackage{dsfont}
\usepackage[font=small,labelfont=bf,justification=justified,format=plain]{caption}
\usepackage{subcaption}

\usepackage{thmtools}
\usepackage{thm-restate}

\definecolor{blueviolet}{rgb}{0.54, 0.17, 0.89}
\definecolor{mycitecolor}{rgb}{0,0.7,0.1}
\usepackage[pdfusetitle]{hyperref}
\hypersetup{
 pdfauthor={Namit Anand},
 pdflang={English},colorlinks=true,linkcolor=blueviolet,citecolor=teal,urlcolor=Periwinkle} 
\usepackage[capitalise]{cleveref}

\graphicspath{{images/}}

\begin{document}
\title{Scrambling and operator entanglement in local non-Hermitian quantum systems}

\author{Brian Barch}
\email[Equal contribution.]{}
\affiliation{Department of Physics and Astronomy, and Center for Quantum Information Science and Technology, University of Southern California, Los Angeles, California 90089-0484, USA}

\author{Namit Anand}
\email[Equal contribution.]{}
\affiliation{Quantum Artificial Intelligence Laboratory (QuAIL), NASA Ames Research Center, Moffett Field, CA, 94035, USA}
\affiliation{KBR, Inc., 601 Jefferson St., Houston, TX 77002, USA}

\author{Jeffrey Marshall}
\affiliation{Quantum Artificial Intelligence Laboratory (QuAIL), NASA Ames Research Center, Moffett Field, CA, 94035, USA}
\affiliation{USRA Research Institute for Advanced Computer Science (RIACS), Mountain View, CA, 94043, USA}

\author{Eleanor Rieffel}
\affiliation{Quantum Artificial Intelligence Laboratory (QuAIL), NASA Ames Research Center, Moffett Field, CA, 94035, USA}

\author{Paolo Zanardi}
\affiliation{Department of Physics and Astronomy, and Center for Quantum Information Science and Technology, University of Southern California, Los Angeles, California 90089-0484, USA}
\date{\today}

\begin{abstract}
The breakdown of Lieb-Robinson bounds in local, non-Hermitian quantum systems opens up the possibility for a rich landscape of quantum many-body phenomenology. We elucidate this by studying information scrambling and quantum chaos in non-Hermitian variants of paradigmatic local quantum spin-chain models. We utilize a mixture of exact diagonalization and tensor network techniques for our numerical results and focus on three dynamical quantities: (i) out-of-time-ordered correlators (OTOCs), (ii) operator entanglement of the dynamics, and (iii) entanglement growth following a quench from product initial states. We show that while OTOCs fail to capture information scrambling in a simple, local, non-Hermitian transverse-field Ising model, the closely related operator entanglement is a robust measure of dynamical properties of interest. Moreover, we show that the short-time growth of operator entanglement can generically detect ``entanglement phase transitions'' in these systems while its long-time average is shown to be a reliable indicator of quantum chaos and entanglement phases. This allows us to extend operator entanglement based diagnostics from previous works on closed and open quantum systems, to the new arena of monitored quantum dynamics. Finally, we remark on the efficacy of these dynamical quantities in detecting integrability/chaos in the presence of continuous monitoring.
\end{abstract}
\maketitle

\section{Introduction}
\label{sec:introduction}
While simply an application of standard quantum mechanics, non-Hermitian physics remains relatively understudied, offering an exciting frontier beyond traditional quantum theory investigations \cite{ashida_non-Hermitian_2020}. As an example, measurement-induced phase transitions in a continuously monitored system -- which can be described by an effective non-Hermitian Hamiltonian under certain assumptions -- have been a topic of great interest recently \cite{PhysRevB.98.205136,PhysRevX.9.031009,PhysRevB.99.224307,PhysRevB.100.134306,PhysRevX.10.041020,PhysRevLett.125.030505,PhysRevX.11.011030, agarwal2023recognizing}. These phenomena have inspired the reexamination of several fundamental results in quantum many-body theory, for example, gap closing across a quantum phase transition \cite{matsumoto_continuous_2020}, classification of topological phases \cite{PhysRevX.9.041015}, bulk-boundary correspondence \cite{PhysRevLett.116.133903}, and many-body localization \cite{hamazaki2019,mak2023statics}, among others. In a similar spirit, it is worth revisiting the interplay of non-Hermitian physics and quantum chaos, which has a long and rich history \cite{Efetov1997,Chalker1997,Fyodorov1997,fyodorov1998universality,Fyodorov2003}; the reader is redirected to Refs. \cite{dalessio_quantum_2016,borgonovi_quantum_2016} for recent reviews on quantum chaos and thermalization. In particular, we would like to focus on local non-Hermitian Hamiltonians and their quantum chaotic properties (as opposed to that of non-Hermitian random matrix ensembles which are highly nonlocal). Building upon the standard random matrix theory classification of Hamiltonians \cite{guhr_random-matrix_1998}, non-Hermitian Hamiltonians are now described by a more general universality class \cite{PhysRevResearch.2.023286,PhysRevLett.123.254101,sa_complex_2020}. Moreover, \textit{complex spacing ratios} \cite{sa_complex_2020} were recently introduced to distinguish the spectral statistics of integrable-vs-chaotic non-Hermitian systems. This allows for generalizing the famous \textit{level-spacing statistics} criteria used to distinguish (Hermitian) integrable and chaotic Hamiltonian systems \cite{haake_quantum_2010}. The usual orthogonality of (nondegenerate) Hamiltonian eigenstates is now replaced by a \textit{biorthogonality} relation \cite{brody_biorthogonal_2013}. These distinctions in the non-Hermitian case require a reinvestigation of conventional wisdom in quantum many-body systems, especially in the presence of open-system effects, see e.g., Refs. \cite{li_spectral_2021,PhysRevResearch.4.L042004} for some recent works.

A key nontrivial aspect of non-Hermitian Hamiltonians is that the Lieb-Robinson (LR) bound \cite{liebFiniteGroupVelocity1972, hastingsSpectralGapExponential2006, bravyi_lieb-robinson_2006} can, in general, break down for these systems \cite{PhysRevLett.120.185301,matsumoto_continuous_2020}, leading to nonlocal growth of operators under local Hamiltonians. LR bounds determine a finite speed of operator growth in non-relativistic quantum systems (with a tensor product structure), evolving under local Hamiltonians \cite{liebFiniteGroupVelocity1972}. The bounds have proven to be fundamental in proving a number of key results in quantum many-body theory such as the exponential decay of correlations \cite{hastingsSpectralGapExponential2006,nachtergaele2006lieb}, the Lieb-Schultz-Mattis theorem in higher dimensions \cite{hastings2004lieb,nachtergaele2007multi}, and generation of correlations and topological order \cite{bravyi_lieb-robinson_2006}. While there are a plethora of ways in which LR bounds influence the landscape of many-body phenomenology, our focus is on their interplay with quantum chaos as traditionally quantified in Hermitian Hamiltonian systems \cite{haake_quantum_2010}.

The primary focus of this work is on quantum lattice models, though the results easily generalize to broader settings. Since local, non-Hermitian lattice models generically violate the LR bound, a natural question that arises is whether these systems can host a notion of ``strong'' quantum chaos \cite{kukuljan_weak_2017-1}. In this regard, we construct two non-Hermitian extensions of the transverse field Ising model, one with an added local imaginary magnetic field (equivalent to weak measurement with postselection), and one where non-Hermiticity acts via similarity transform, and thus preserves the spectrum. The latter model is a nontrivial quantum chaotic Hamiltonian: a locally interacting spin chain, whose spectrum is chaotic while its eigenstates can transition from volume-law to area-law by controlling the measurement rate (see \cref{fig:TFIM-chaos}). As a result, for this model, the spectral quantities always satisfy chaotic features while eigenstate properties do not. A Hermitian lattice model with a ``chaotic spectrum'' and ``integrable eigenstates'' was previously introduced by a subset of the authors in a recent work \cite{anand_brotocs_2021_published}; however in the current model, the nontriviality originates from continuous monitoring rather than by construction.

The chaoticity of these non-Hermitian TFIMs is studied as a function of system size and degree of nonhermiticity, using out-of-time-ordered correlators (OTOCs) and operator entanglement of the time evolution operator. To the best of our knowledge, a consistent way to define time evolution of operators (i.e., the Heisenberg picture) is nonexistent (it is easy to show that it has inconsistencies with the Schrodinger picture). As a result, when studying the OTOC, we introduce a new definition that focuses on the evolution of states instead. Similarly, an alternative based on a slightly modified Heisenberg evolution is described in the Appendix. We find the operator entanglement to be a more robust measure of chaoticity than OTOCs as we will detail in \cref{sec:otocs-vs-operatorentanglement-non-Hermitian}. In particular, we show that the growth with system size of the operator entanglement long-time average (LTA) can distinguish chaotic and integrable non-Hermitian models, and analytically derive approximation to the LTA that captures this scaling behavior. For the first of the two non-Hermitian TFIMs, we find a non-monotonic relationship between measurement strength and operator entanglement LTA, which we study in terms of the Hamiltonian's now complex spectrum.

\section{Background}
  
\subsection{Out-of-time-ordered correlators}

Let \(\mathcal{H} \cong \mathbb{C}^{d}\) be a finite, \(d\)-dimensional Hilbert space and
\(\mathcal{L}(\mathcal{H})\) denote the space of linear operators on
\(\mathcal{H}\). We will endow \(\mathcal{H}\) with a tensor
product/lattice structure, e.g., \(\mathcal{H} \cong \left(
  \mathbb{C}^{2} \right)^{\otimes L}\) with \(L\) the system size and \(d = 2^{L}\). For quantum evolution generated by a Hamiltonian,
scrambling can be quantified by considering two, typically local, operators \(V,W
\in \mathcal{L}(\mathcal{H})\). Let \(W_{t}\equiv U^{\dagger}_{t} W
U_{t}\) denote the time evolution generated by the dynamical unitary, \(U_{t} \equiv e^{-iHt}\) in the Heisenberg picture. We consider the norm of
the commutator between the static operator $V$ and the dynamical one, $W_t$, i.e., \cite{larkin_quasiclassical_1969, kitaev_simple_2015,swingle_unscrambling_2018,xu_swingle_tutorial_2022}
\begin{align}
\begin{split}
\label{eq:OTOC-Hermitian}
  C_{V,W}(t) &\equiv \frac{1}{2d} \left\Vert \left[ W_{t},V \right]  \right\Vert_{2}^{2}\\
  &=  \frac{1}{2d} \operatorname{Tr}\left[ \left[ W_{t},V \right]^{\dagger} \left[ W_{t},V \right]   \right]\\
  &= \frac{1}{d}\left[ \left\Vert V W_t  \right\Vert_2^2 - \mathrm{Re} \operatorname{Tr}\left[ W^{\dagger}_{t} V^{\dagger} W_{t} V \right] \right]
\end{split}
\end{align}
where \(\left\Vert X \right\Vert_{2}^{2}  \equiv \operatorname{Tr}\left[
  X^{\dagger}X \right]\) is the Hilbert-Schmidt operator norm. If $V,W$ are further assumed to be unitary, then the expression above can be simplified to,
\begin{align}
\label{eq:unitary-OTOC}
C_{V,W}(t) = 1 - \frac{1}{d} \mathrm{Re} \operatorname{Tr}\left[ W^{\dagger}_{t} V^{\dagger} W_{t} V \right].
\end{align}
The quantity,
\begin{align*}
  F_{V,W}(t)\equiv \frac{1}{d} \mathrm{Re} \operatorname{Tr}\left[ W^{\dagger}_{t} V^{\dagger} W_{t} V \right]
\end{align*}
is the so-called four-point out-of-time-ordered correlator (OTOC) \cite{larkin_quasiclassical_1969, kitaev_simple_2015,swingle_unscrambling_2018,xu_swingle_tutorial_2022}. Note that the norm of the
commutator and the OTOC introduced here are both defined for the
infinite-temperature case, which will be our focus. Moreover, since \(C_{V,W}(t)\) and
\(F_{V,W}(t)\) are related to each other via a simple affine relation,
we will interchangeably refer to them as the OTOC.

OTOCs have been applied to study a variety of many-body phenomena, ranging from quantum phase transitions~\cite{PhysRevLett.123.140602,Zamani2022OTOCs} to many-body localization~\cite{Huang_2016,Fan_2017,chen2016universal,vonKeyserlingk:2017dyr,yunger_halpern_quasiprobability_2018}. Moreover, connections with dynamical quantities have also been discovered such as the Loschmidt Echo \cite{yan_information_2020}, operator entanglement and local entropy production \cite{styliaris_information_2021, zanardi_information_2021-1}, quantum coherence \cite{anand_quantum_2020_published}, quasiprobabilities \cite{yunger_halpern_quasiprobability_2018}, entropic uncertainty relations \cite{halpern_entropic_2019}, and even information-theoretic hardness of learning quantum dynamical features, see, e.g., Ref.~\cite{cotler_otoc_hardness}.

OTOCs have, by now, also been measured in a number of state-of-the-art experimental setups, e.g., using superconducting qubits ~\cite{mi2021information,braumuller2021probing}, nuclear magnetic resonance~\cite{Wei_2018,li_measuring_2017,nie2019detecting,Nie_2020}, and ion-trap quantum simulators~\cite{garttner_measuring_2017,Joshi_2020}, among others~\cite{Meier_2019,Chen_2020}.

We briefly summarize the key results relating averaged OTOCs to
dynamical quantities such as operator entanglement and entangling
power. The bulk of these results were obtained in
Ref.~\cite{styliaris_information_2021}, which was then generalized to open
quantum systems \cite{zanardi_information_2021-1}, finite temperature
\cite{anand_brotocs_2021_published}, and general observable algebras for closed
\cite{zanardi_quantum_2021_published} and open systems
\cite{andreadakis_scrambling_2022_published}. Given a bipartite Hilbert space,
\(\mathcal{H}_{AB} = \mathcal{H}_{A} \otimes \mathcal{H}_{B} \cong
\mathbb{C}^{d_{A}} \otimes \mathbb{C}^{d_{B}}\), let
\(\mathcal{U}(\mathcal{H}_{A})\) denote the unitary
group over \(\mathcal{H}_{A}\) (similarly for subsystem \(B\)). We will from now
on focus on OTOCs of the form, \(C_{V_{A},W_{B}}(t)\), where \(V_{A}
\equiv V \otimes \mathbb{I}_{B}\) and \(W_{B} \equiv \mathbb{I}_{A}
\otimes W\) are local operators with support on subsystems \(A,B\),
respectively. Note that such an OTOC vanishes at \(t = 0\) since \(
\left[ V_{A},W_{B} \right] =0\) and grows over time as the support of the Heisenberg evolved operator,
\((W_{B})_t\) grows and starts to overlap with subsystem \(A\). We are
now ready to define the ``averaged bipartite OTOC,''
\cite{styliaris_information_2021},
\begin{align}
  \label{eq:bipartite-otoc}
G(t) \equiv \mathbb{E}_{V_{A},W_{B}} \left[ C_{V_{A},W_{B}}(t) \right], 
\end{align}
where \(\mathbb{E}_{V_{A}} \left[...\right]\) denotes Haar
averaging over
\(\mathcal{U}(\mathcal{H}_{A})\) (and similarly for
\(\mathbb{E}_{W_{B}}\)). That is, starting from the OTOC,
\(C_{V_{A},W_{B}}(t)\), we average over the local unitaries \(V_{A},W_{B}\) and therefore, the OTOC is now only a function of the
dynamical unitary \(U_{t}\) (and the choice of bipartition
\(A|B\)). 
Quite surprisingly, and as first noted in Ref.~\cite{styliaris_information_2021}, this allows us to make a connection with the operator entanglement of the time evolution operator, \(U_{t}\), which we will now introduce.

\subsection{Operator entanglement and its connection to OTOCs}
Given a linear operator, \(X \in \mathcal{L}(\mathcal{H}_{A} \otimes
\mathcal{H}_{B})\), we can define an \textit{operator Schmidt decomposition} \cite{zanardi_entanglement_2001-1,lupo2008bipartite,aniello2009relation}. Formally, given a bipartite operator \(X \in \mathcal{L}(\mathcal{H}_{A} \otimes \mathcal{H}_{B})\), there exist orthogonal bases \(\{ U_{j} \}_{j=1}^{d_{A}^{2}}\) and \(\{ W_{j} \}_{j=1}^{d_{B}^{2}}\) for \(\mathcal{L}(\mathcal{H}_{A}), \mathcal{L}(\mathcal{H}_{B})\), respectively, such that \(\left\langle U_{j},U_{k} \right\rangle = \text{Tr}[ U_j^\dag U_k] = d_{A} \delta_{jk}\) and \(\left\langle W_{j}, W_{k} \right\rangle = d_{B} \delta_{jk}\). Moreover, there exist $\lambda_j\geq0$ such that
\begin{align}
X = \sum\limits_{j=1}^{\tilde{r}} \sqrt{\lambda_{j}} U_{j} \otimes W_{j}.
\end{align}
The coefficients \(\{ \lambda_{j} \}_{j}\) are
called the operator Schmidt coefficients and \(\tilde{r} = \min \{
d_{A}^{2},d_{B}^{2} \}\) is the operator Schmidt rank. In fact, the operator entanglement of a unitary introduced in Ref.~\cite{zanardi_entanglement_2001-1} is exactly the linear entropy of the probability vector \(\vec{p} = (\lambda_{1}, \lambda_{2}, \cdots, \lambda_{\tilde{r}})\) arising from the operator Schmidt coefficients. A key result obtained in Ref.~\cite{zanardi_entanglement_2001-1} was that the operator entanglement of a unitary operator $U$ can be equivalently expressed as
\begin{align}
\label{eq:lin_herm_opent}
E_{\mathrm{op}}^\text{lin} \left( U \right) = 1 - \frac{1}{d^{2}} \operatorname{Tr}\left[ \mathbb{S}_{AA'} U^{\otimes 2} \mathbb{S}_{AA'} U^{\dagger \otimes 2}\right],
\end{align}
where $\mathbb{S}_{AA'}$ is the partial SWAP operator, acting on the $A$ subsystem and its copy $A'$.

The key result of Ref.~\cite{styliaris_information_2021} is that
\begin{align*}
G(t) = \mathbb{E}_{V_{A},W_{B}} \left[ C_{V_{A},W_{B}}(t) \right] = E_{\mathrm{op}}^\text{lin}(U_{t}).
\end{align*}
That is, the OTOC when averaged over local, Haar random unitaries is
exactly equal to the operator entanglement of the dynamical unitary
\(U_{t}\). Moreover, typicality ensures that in higher dimensional systems, a random instance of \(C_{V_{A},W_{B}}(t)\) is exponentially close to
\(E_{\mathrm{op}}^\text{lin}(U_{t})\); see, e.g., Proposition 3 of
Ref.~\cite{styliaris_information_2021}. As a result, intuitively,
studying \textit{almost any} nonlocal OTOC \(C_{V_{A},W_{B}}(t)\),
with \(V_{A},W_{B}\) being nonlocal operators, is
equivalent to studying the operator entanglement of \(U_{t}\). As we will see, this requires reexamination when addressing nonunitary time evolution channels.

Before proceeding further it is worth clarifying the distinction between operator entanglement versus ``local operator entanglement,'' both of which have been widely studied in quantum chaotic systems. Local operator entanglement is simply the entanglement of a Heisenberg evolved \textit{local} operator. As an example consider a Pauli operator on site \(j\), say \(\sigma^x_j\). Local operator entanglement is the entanglement of the following operator: \(\sigma^x_j(t) := e^{iHt} \sigma^x_j e^{-iHt}\). For quantum chaotic systems this generically grows linearly in time and has been studied for example in Refs. \cite{Prosen2007,Piorn2009,dubail_entanglement_2017,Alba2019,Hartmann2009,Muth2011,nidari2008,Cheryne2018}. In contrast, our work is focused on ``global operator entanglement'', namely, the operator entanglement of the time-evolution operator, \(U(t) = e^{-iHt}\) itself. This has been studied for example in Refs. \cite{hosur_chaos_2016,Jonnadula2017,Lensky2019,Pal2018,styliaris_information_2021,zanardi_information_2021-1,anand_brotocs_2021_published,zanardi_quantum_2021_published,andreadakis_scrambling_2022_published}. We refer the reader to \cite{dowling_scrambling_2023} for the interplay between scrambling and local operator entanglement, and to \cite{styliaris_information_2021} for the interplay between (global) operator entanglement and scrambling.

\subsection{Quantum chaos}
The ability to classify quantum systems simply from their ergodicity (or lackthereof) has been a fundamental direction of research in many-body physics for the last few decades \cite{dalessio_quantum_2016,borgonovi_quantum_2016}. In this regard, the developments in quantum chaos have been quite successful, such as in extending the classification of integrable and chaotic models, solely using their spectral statistics, to the quantum domain. As is well-established now, the spectral statistics of quantum chaotic systems are described by random matrix theory, depending, e.g., on the symmetry class for Hamiltonian systems \cite{mehta_random_2004,livan_introduction_2018}. On the other hand, integrable and (many-body) localized models generically feature Poisson statistics \cite{nandkishoreManyBodyLocalizationThermalization2015,abanin_colloquium_2019,alet2018many}. The success of classifying ergodic properties using spectral quantities has also inspired the search for other criteria, such as properties of the eigenstate entanglement \cite{huang2019universal,Kumari2022eigenstate,RigolEE2017,RigolEE2017,beugeling2015global,PhysRevLett.119.020601,PhysRevB.99.075123,PhysRevE.100.022131,PhysRevLett.125.180604}, dynamical quantities such as Loschmidt echo \cite{Goussev:2012,Gorin2006LEreview}, OTOCs \cite{https://doi.org/10.48550/arxiv.2209.07965, dowling_scrambling_2023}, etc. 

In this regard, perhaps the most distinct bifurcation is exemplified by the contrast between quantum chaotic Hamiltonians and Anderson (or many-body) localized models. The former typically show thermalization of the expectation values of \textit{local} observables, have few (or nonextensive) conserved quantities, and generally have volume-law eigenstate entanglement (in the thermodynamic limit) \cite{dalessio_quantum_2016,borgonovi_quantum_2016}. In contrast, localized models generally \textit{escape} thermalization, have an extensive number of (quasi-)local integrals of motion, and show area-law eigenstate entanglement with exponentially small violations \cite{nandkishoreManyBodyLocalizationThermalization2015,abanin_colloquium_2019,alet2018many}.

While this classification has been thoroughly examined for Hermitian systems (or unitary dynamics), the generalization to non-Hermitian systems (e.g., non-unitary dynamics) is just beginning to be explored. Dissipative systems are no longer classified within the previous framework, rather, it has been suggested that integrable systems are expected to follow Poisson level statistics while chaotic systems follow the statistics of the \textit{Ginibre ensemble} \cite{PhysRevResearch.2.023286,PhysRevLett.123.254101,sa_complex_2020,li_spectral_2021}. However, if the spectrum is complex then there are further nontrivialities. All of this has initiated a program to understand the features of quantum chaotic systems in the presence of non-Hermitian system effects. Some approaches include (i) the introduction of complex spacing ratios as a generalization of the universal spectral statistics \cite{sa_complex_2020}, (ii) the introduction of a dissipative form factor \cite{li_spectral_2021}, in analogy to the well-studied and universal spectral form factor in Hermitian systems \cite{berry1985semiclassical,Cotler2017bhrmt,cotler2017chaos}, among others. 

At the same time, there is a novel interest in ``hybrid'' quantum circuits, where one has unitary dynamics with measurements interspersed throughout the evolution \cite{PhysRevB.98.205136,PhysRevX.9.031009,PhysRevB.99.224307,PhysRevB.100.134306,PhysRevX.10.041020,PhysRevLett.125.030505,PhysRevX.11.011030, weinstein2022scrambling}. Furthermore, if one postselects on the condition that there are ``no quantum jumps'' then the effective dynamics can be described by a non-Hermitian Hamiltonian \cite{ashida_non-Hermitian_2020}. If the unitary dynamics is chaotic, then, by tuning the measurement rate, the system can transition from a volume-law entanglement to an area-law entanglement (e.g., in the steady-state entanglement of a initial product state) \cite{PhysRevB.98.205136}. Such transitions have been termed ``measurement-induced entanglement phase transitions''. 
It was found that one can detect the critical point by solely studying the spectral properties of the effective non-Hermitian Hamiltonian, see e.g., \cite{gopalakrishnan_entanglement_2021}. Our isospectral model has the same spectrum but the transition is only in the eigenstate properties and hence cannot be diagnosed with standard spectral measures from the theory of quantum chaos.

\subsection{Non-Hermitian quantum mechanics}
There are myriad contexts for studying non-Hermitian Hamiltonians, but the most common is the study of conditional time evolutions, particularly the no-jump trajectory \cite{ashida_non-Hermitian_2020,brunSimpleModelQuantum2002}. Consider a state $\rho$ evolving under the standard Lindblad equation \cite{lidarLectureNotesTheory2019}

\begin{align*}
    \dot \rho = -i[H,\rho]+\sum_a \left( L_a \rho L_a^\dag - \frac{1}{2}\{L_a^\dag L_a, \rho\} \right).
\end{align*}

The first and third terms correspond to evolution within a single quantum trajectory, while the second term $L_a \rho L_a^\dag$ corresponds to jumps between quantum trajectories, the so-called \textit{quantum jump term}. If we drop this term (i.e., postselect on the evolution conditional on no quantum jumps), the resulting evolution can be described as a pure state evolution under a non-Hermitian effective Hamiltonian 
\begin{align*}
    H_{\mathrm{eff}} = H -\frac{i}{2} \sum_a L_a^\dag L_a.
\end{align*}
The time evolution operator can be written in terms of $H_{\mathrm{eff}}$ as $U_t = e^{-i H_{\mathrm{eff}} t}$, which is no longer unitary.

Note that there are two approaches to studying dynamical properties of non-Hermitian
Hamiltonians: (i) normalization of the
non-unitary evolution or (ii) the metric formalism \cite{ashida_non-Hermitian_2020,mostafazadeh_pseudo-hermiticity_2002}. The former describes a system where the non-Hermitian Hamiltonian represents
the \textit{effective} interaction of the conditional evolution, and is physically applicable. The latter refers to the non-Hermitian Hamiltonian being
\textit{fundamental} and focuses on constructing a new metric for the
Hilbert space, such that the non-Hermitian Hamiltonian is Hermitian
\textit{with respect to} the modified inner product. We will focus on
the former here. This is e.g., the approach in studying measurement-induced phase transitions.

In order for $U_t$ to map states to states, we must normalize its action on pure states $\vert\psi_t\rangle$
\begin{align}
\label{eq:pure-evolution}
    \Vert\psi_t\rangle \equiv \frac{U_t\vert\psi_0\rangle}{\left\Vert U_t\vert\psi_0\rangle\right\Vert}
\end{align}
or trace-normalize its action on mixed states $\rho_t$:
\begin{align}
\label{eq:mixed-evolution}
    \rho_t \equiv \frac{U_t \rho_0 U_t^\dag}{\text{Tr}[U_t \rho_0 U_t^\dag]}
\end{align}
This is equivalent to dividing by the normalizing factor in the Bayes rule for conditional probability distributions, and comes from the fact the total probability of the current conditional trajectory is not constant in time \cite{PhysRevLett.120.185301}. For this reason the divergence of $\rho_t$ when $\text{Tr}[U_t \rho_0 U_t^\dag] = 0$ is no problem conceptually, as the case is unphysical, i.e. occurs with probability zero.

The normalization condition breaks down in the Heisenberg picture of time evolution, making operator-based quantities such as OTOCs and connected correlators require redefinition to remain meaningful \cite{PhysRevLett.120.185301,PhysRevResearch.4.033250}. Even excluding normalization, the time evolution of operators is \textit{nonunital}, i.e., the infinite-temperature state, $\rho = \mathbb{I}/d$ is no longer a fixed-point. This causes, e.g., the seeming generation of nonlocal operator growth from product evolutions:
\begin{align*}
    (U^\dag \otimes V^\dag)(X \otimes I )(U \otimes V) = (U^\dag X U)\otimes (V^\dag V)
\end{align*}

This can be understood in a dynamical form as the breakdown of LR bounds, as described in Ref.~\cite{matsumoto_continuous_2020}. Consider a Hamiltonian $H_{\mathrm{eff}} = \sum_R H_R + i\Gamma_R$ composed of Hermitian $H_R$ and $\Gamma_R$, acting on local regions $R$, and operator $X$ initially localized on region $R_o$. The time evolution of $X$ (excluding normalization) under this Hamiltonian is
\begin{align*}
    \dot X &= i H_{\mathrm{eff}}^\dag X -i X H_{\mathrm{eff}}\\
    &= i\sum_{R\cap R_o \neq \varnothing} [H_R,X]-\sum_R \{\Gamma_R, X\}.
\end{align*}
While $[H_R,X]=0$ for $R\cap R_o = \varnothing$ at $t=0$, the same does not hold for $\{\Gamma_R,X\}$, which can in general cause nonlocal growth of $X$. Including normalization yields a similar equation and does not in general return locality.

\subsection{Non-Hermitian Hamiltonians}
When $H_{\mathrm{eff}}$ is nondegenerate, it may be diagonalized as 
\begin{align}
\label{eq:diagonal}
    H_{\mathrm{eff}} = \sum_i \lambda_i \vert r_i \rangle \langle l_i \vert
\end{align}
for $\lambda_i$ the (in general complex) eigenvalues, and $\vert r_i\rangle$ and $\langle l_i \vert$ the right and left eigenvectors, respectively. We can always bi-orthonormalize the eigenvectors such that $\langle l_i\vert r_j \rangle = \delta_{ij}$, and may additionally normalize $\langle r_i \vert r_i\rangle = 1$, but cannot do the same for $\langle l_i \vert$. Given this normalization choice, we can write the time evolution operator in terms of the same components as
\begin{align}
    U_t = \sum_i e^{-it \lambda_i} \vert r_i \rangle \langle l_i \vert.
\end{align}
Notice that while real $\lambda_i$ generate periodic behavior, complex $\lambda_i$ cause exponential growth or decay in their respective eigenspaces. When such $\lambda_i$ exist, $U_t$ is said to be purifying, as once normalization is included the time evolution at large $t$ will effectively project into the fastest growing (or slowest decaying) subspace. We refer to this as the long-time eigenspace, and its dimension is equal to the number of eigenvalues that share the maximal imaginary component. 

Non-Hermitian Hamiltonians are said to be pseudohermitian if they satisfy the equation $H_\text{eff}^\dag \eta = \eta H_\text{eff}$ for some (non-unique) Hermitian $\eta$, which is dubbed the metric as it defines a modified inner product under which $H_\text{eff}$ is Hermitian \cite{Bender_1998_real_spectra, Bender_2002_complex_extension, mostafazadeh_pseudo-hermiticity_2002}. In this case the eigenvaleus of $H_\text{eff}$ come in complex conjugate pairs. If $\eta$ is further taken to be positive definite, $H_\text{eff}$ is referred to as quasihermitian and has real spectrum. While nonunitary $U_t$ no longer generates standard norm-preserving (i.e. spherical) Hilbert space rotations, it now generates evolutions that preserve the norm with respect to $\eta$, which can be seen as elliptical (quasihermitian) or hyperbolic (pseudohermitian) rotations.

The spectra of Hamiltonians such as in Ref.~\cite{gopalakrishnan_entanglement_2021} can transition from real to complex spectrum as the degree of nonhermiticity is increased. At the border between the two regimes is an exceptional point, where eigenvalues and eigenvectors coalesce. In this case (and for general nondiagnalizable $H_{\mathrm{eff}}$) \cref{eq:diagonal} must be replaced by the Jordan normal form. This leads to additional polynomial terms in $U_t$ and generally unique behavior. The physical significance of the exceptional point is unclear, since e.g. in Refs.~\cite{agarwal2022detecting,lakkaraju2021detecting} it can be related to a factorization surface, while in Ref.~\cite{PhysRevResearch.4.033250} it corresponds to conditioning on a trajectory which has probability zero of occurring.

\section{Quantum spin-chain models}
We wish to study the dynamical features of local, non-Hermitian
quantum spin-chain models. As a paradigmatic quantum spin-chain model, we study the transverse-field Ising model (TFIM)

\begin{align}
\label{eq:TFIM-hermitian}
H_{\mathrm{TFIM}}= J\sum\limits_{j=1}^{L-1} \sigma_{j}^{z} \sigma_{j+1}^{z} + g \sum\limits_{j=1}^{L} \sigma_{j}^{x} + h \sum\limits_{j=1}^{L} \sigma_{j}^{z},
\end{align}

which can host an integrable, chaotic, and a localized phase. Here \(\sigma_{j}^{\alpha},\ \alpha \in \{ x,y,z \}\) are the Pauli matrices, and \(g,h\) denotes the strength of the transverse field and the local field, respectively. The TFIM Hamiltonian is integrable for either \(h=0\) or \(g=0\) and nonintegrable when both \(g,h\) are nonzero. We take $J=0.95$ and consider as the integrable point, \(g=1,\ h=0\) and the nonintegrable point \(g=1,\ h=0.5\). At the integrable point, the TFIM can be mapped onto free fermions via the Jordan-Wigner transformation and is ``highly integrable'' in this sense. At the nonintegrable point, the model is quantum chaotic in the sense of random matrix spectral statistics \cite{PhysRevLett.106.050405,PhysRevLett.111.127205} and volume-law entanglement of eigenstates \cite{wolfAreaLawsQuantum2008}. We also consider as the classical point $g=0,\ h=0.5$, where $H_\text{TFIM}$ is diagonal in the computational basis.

We consider two non-Hermitian extensions of the TFIM. The first, termed the ``measurement-induced TFIM,'' was considered in Refs. \cite{gopalakrishnan_entanglement_2021,Biella2021manybodyquantumzeno} where its spectral and eigenstate properties were studied to characterize its phase transition. The model is a TFIM with an additional imaginary field along the \(y\)-axis, described by the Hamiltonian,
\begin{align}
\label{eq:meas-TFIM}
H_\mathrm{M} \equiv H_\mathrm{TFIM} + i \gamma \sum\limits_{j=1}^{L} \sigma_{j}^{y}
\end{align}
where \(\gamma\) denotes the strength of the imaginary field. This model can be generated by application of $H_\mathrm{TFIM}$ combined with continuous weak measurement of local $y$ spins on all qubits, postselected to always yield measurement results of $+1$. In this sense the nonhermiticity parameter $\gamma$ is a form of measurement strength. At $\gamma = g$, the $\sigma^x$ and $\sigma^y$ terms combine into $\sigma^+$ making $H_\mathrm{M}$ upper triangular in the $z$-basis and highly degenerate. At separate $\gamma > g$, $H_\mathrm{M}$ undergoes multiple exceptional points as the spectrum becomes complex. In the chaotic case, this generates an ``entanglement phase transition'' from the volume-law phase to an area-law phase  \cite{gopalakrishnan_entanglement_2021,Biella2021manybodyquantumzeno}.

The second, closely related, model studied is termed the "isospectral TFIM" and is motivated by the non-Hermitian model in Ref.~\cite{matsumoto_continuous_2020}. Starting with the Hermitian TFIM in \cref{eq:TFIM-hermitian}, perform the similarity transform
\begin{align}
\label{eq:isospectral-TFIM}
H_\mathrm{I} \equiv S(\beta) H_\mathrm{TFIM} S(\beta)^{-1},
\end{align}
where
\begin{align}
S(\beta) \equiv \exp \left[ \frac{\beta}{2}  \sum\limits_{j=1}^{L} \sigma_{j}^{z}\right] = \prod_{j=1}^{L} \exp \left[ \frac{\beta}{2} \sigma_{j}^{z} \right]
\end{align}
is a case of the Dyson map \cite{Fring2016, Dyson1956}. Note that $S(\beta)$ acts nontrivially only on $\sigma^x$ terms in $H_\mathrm{TFIM}$, so one can calculate
\begin{align*}
& H_\mathrm{I} \\
&= J \sum\limits_{j=1}^{L-1} \sigma_{j}^{z} \sigma_{j+1}^{z} + g \cosh(\beta) \sum\limits_{j=1}^{L} \sigma_{j}^{x} + i g \sinh(\beta) \sum\limits_{j=1}^{L} \sigma_{j}^{y} \\
&= H'_{\mathrm{TFIM}} + i g \sinh(\beta) \sum\limits_{j=1}^{L} \sigma_{j}^{y}.
\end{align*}
That is, it is equivalent to a measurement-induced TFIM with a altered transverse-field $g' \equiv g \cosh(\beta)$ and imaginary field strength $\gamma =  g \sinh(\beta)$. Note that for the classical TFIM with $g=0$, $H_\mathrm{I}$ remains Hermitian, making this specific case uninteresting.

Perhaps a more interesting way to rewrite the model is as follows -- it makes apparent why the Hamiltonian never breaks \(PT\)-symmetry, but instead has a smooth measurement-induced entanglement phase transition. This is because the model is equivalent to 
\begin{align}
\label{eq:tfim-isospectral}
H_\mathrm{I} = J \sum_{j=1}^{L-1} \sigma_{j}^{z} \sigma_{j+1}^{z} + h \sum\limits_{j=1}^{L} \sigma_{j}^{z} + g \sum\limits_{j=1}^{L} \left( e^{\beta} \sigma_{j}^{+} + e^{-\beta} \sigma_{j}^{-} \right),
\end{align}
where \(\sigma_{j}^{\pm} \equiv \frac{1}{2} \left( \sigma_{j}^{x} \pm i \sigma_{j}^{y} \right)\). Clearly, at \(\beta \ll 1\) the model is well approximated by the Hermitian TFIM while at \(\beta \gg 1\) the model is dominated by the \(\sigma^{+}\) terms and becomes noninteracting, at which point the eigenstates of this model are simply determined by the eigenstates of \(\sigma^{+}\).

Notice that there is a unique state stationary for the evolution under $H_I$ regardless of the choice of parameters in $H_\text{TFIM}$. This is analogous to the fact that the maximally mixed state, $\mathbb{I}/d$, is a stationary state (or fixed point) for unitary evolutions. In Appendix~\ref{app:steadystate} we show that the following mixed state is \textit{stationary},
\begin{align}
    \rho_{ss} = S(\beta)^2/\text{Tr}[S(\beta)^2].
    \label{eq:stationary}
\end{align}
Moreover, it has a purity \(\text{Tr}[\rho_{ss}^2]=2^{-L}\left[1-\text{tanh}(\beta)^2\right]^L\).

\prlsection{Remark on the choice of spin-chain models} The two models considered above are both of one-dimensional systems whose purely unitary dynamics is scrambling and they undergo an entanglement phase transition in the presence of (disentangling) measurements; and therefore fall into the same universality class, see Ref. \cite{PhysRevX.9.031009} for an excellent discussion. Moreover, our specific choice is motivated from the following criteria: (i) we want to focus on a local many-body chaotic Hamiltonian (as opposed to say a nonlocal random matrix ensemble), (ii) we want to focus on models well-studied in the literature, e.g., Refs. \cite{gopalakrishnan_entanglement_2021, turkeshi_measurement-induced_2021} study similar models, and (iii) they can illustrate the eigenstate vs. joint eigenstate and spectral transition that we would like to elucidate (namely the ``isospectral TFIM'' in contrast to the ``measurement-induced TFIM''). With reference to these criteria, our models serve as both illustrative and paradigmatic examples.

\section{OTOCs vs operator entanglement for non-Hermitian Hamiltonians}
\label{sec:otocs-vs-operatorentanglement-non-Hermitian}

As previously mentioned, OTOCs are a well studied tool for diagnosing many-body behavior, such as the LR bound. The traditional OTOC is defined in terms of Heisenberg evolution of operators, and does not immediately generalize to non-Hermitian models. In order to generalize the OTOC, we must either redefine the OTOC in terms of an evolution of states, or define a modified Heisenberg evolution of operators. We take the former approach here as it is closer to the literature of measurement-induced phase transitions and non-Hermitian Hamiltonians. The latter approach is discussed in Appendix~\ref{app:OTOC}.

\begin{figure*}[!ht]
\raggedright
\begin{subfigure}{.245\textwidth}
\includegraphics[width=0.9\linewidth]{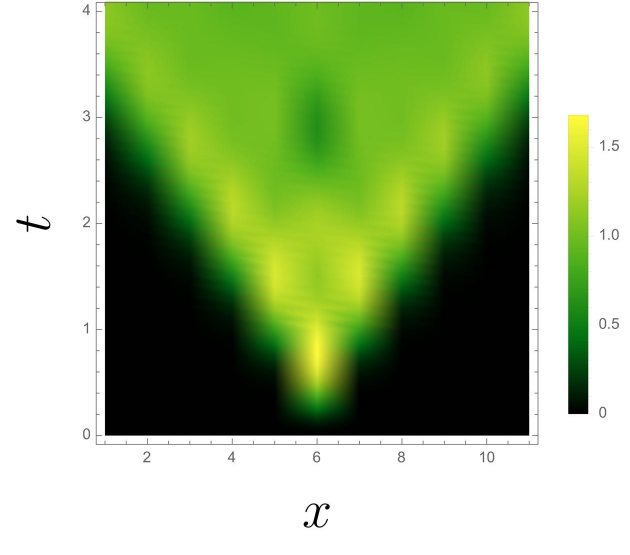}
\caption{\(\beta = 0\)}
\end{subfigure}
\begin{subfigure}{.245\textwidth}
\includegraphics[width=0.9\linewidth]{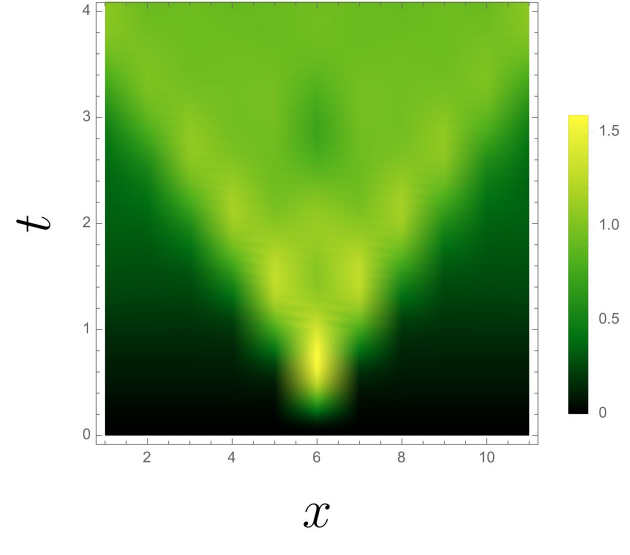}
\caption{\(\beta = 0.25\)}
\end{subfigure}
\begin{subfigure}{.245\textwidth}
\includegraphics[width=0.9\linewidth]{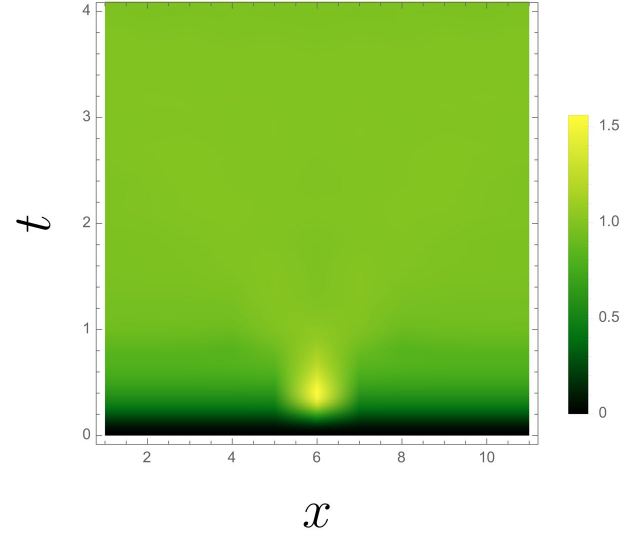}
\caption{\(\beta = 1\)}
\end{subfigure}
\begin{subfigure}{.245\textwidth}
\includegraphics[width=0.9\linewidth]{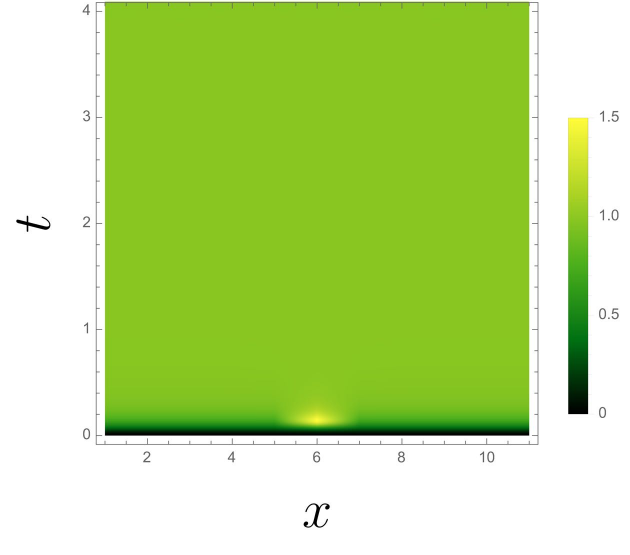}
\caption{\(\beta = 2\)}
\end{subfigure}%
\caption{Breakdown of the lightcone spreading of OTOCs (see \cref{eq:OTOC-namit}) in the isospectral TFIM. $C_{V,W}(t)$ is plotted for \(L=11\) spins, where the operators are \(V = W = \sigma_z\), $V$ acts on the fifth qubit and $W$ is swept across the 1D chain. As the strength of non-Hermiticity is increased from \(\beta = 0\) (Hermitian model) to \(\beta = 0.25\), the lightcone becomes ``fuzzy''. As we keep increasing the \(\beta\), at \(\beta=1\) the lightcone almost vanishes and at \(\beta = 2\), the OTOCs saturate immediately. This is in stark contrast to how the operator entanglement in this system behaves; which shows a smooth transition from a volume-law phase to an area-law phase as \(\beta\) increases (cf. \cref{fig:TFIM-chaos}).}
\label{fig:lieb-robinson-9qubits-isospectral}
\end{figure*}

\prlsection{Defining a normalized OTOC} Given an orthonormal basis $\{\vert j \rangle \}_{j=1}^d$ for the Hilbert space \(\mathcal{H} \cong \mathbb{C}^{d}\), we can write the traditional OTOC in \cref{eq:OTOC-Hermitian} as

\begin{align}
    C_{V,W}(t) = \frac{1}{d} \sum\limits_{j=1}^{d} \left[\left\Vert V W_t |  j \rangle \right\Vert_2^2- \mathrm{Re} \langle j | W^{\dagger}_{t} V^{\dagger} W_{t} V |  j \rangle \right].
\end{align}

We generalize this to the non-Hermitian case by applying the normalized pure state evolution in \cref{eq:pure-evolution} to $W_t\vert j\rangle$ and $W_t V \vert j\rangle$. This gives the normalized OTOC

\begin{align}
\hat C_{V,W}(t) \equiv \frac{1}{d} \sum\limits_{j=1}^{d} \left[ \frac{\left\Vert V W_{t} |  j \rangle\right\Vert_2^2}{\left\Vert W_t | j \rangle  \right\Vert_{2}^{2}} - \mathrm{Re} \frac{\langle j | W^{\dagger}_{t} V^{\dagger} W_{t} V |  j \rangle}{\left\Vert W_{t} | j \rangle \right\Vert_{}^{} \left\Vert W_{t} V | j \rangle \right\Vert } \right].
\end{align}

The normalization breaks basis invariance of the trace, introducing basis dependency into the OTOC. For the calculations performed herein, the computational basis is used. Like in the Hermitian case, if $V$ is unitary the first term is simply equal to one. In this case, we have
\begin{align}
\label{eq:OTOC-namit}
\hat C_{V,W}(t) = 1 - \frac{1}{d}  \sum\limits_{j=1}^{d} \frac{\mathrm{Re} \langle j | W^{\dagger}_{t} V^{\dagger} W_{t} V |  j \rangle}{\left\Vert W_{t} | j \rangle \right\Vert_{}^{} \left\Vert W_{t} V | j \rangle \right\Vert_{}^{} }.
\end{align}

This normalized OTOC is useful, for example, for the study of LR bound violations as in \cref{fig:lieb-robinson-9qubits-isospectral}, as the normalization prevents exponential growth and recovers the bound $0\leq \hat C_{V,W}(t) \leq 2$ as in the Hermitian case.

Though we would like to be able to take an analytic bipartite Haar average 
as in Ref.~\cite{styliaris_information_2021}, the normalization term makes this elusive. Furthermore, as shown in Appendix~\ref{app:otoc-convergence}, numeric averages of the bipartite OTOC converge to a quantity distinct from the operator entanglement, indicating the connection no longer holds for non-Hermitian systems.

\subsection{Operator entanglement and measurement-induced phase transitions}
Motivated by the connection between bipartite OTOCs and operator entanglement for unitary dynamics, we propose studying the dynamical features of non-Hermitian evolutions by \textit{directly} studying the operator entanglement of the time evolution operator, which is already defined in a form that applies to non-unitary time evolutions. For an arbitrary linear operator \(X\), the 2-Renyi operator entanglement is defined, based on the form in Ref.~\cite{zanardi_entanglement_2001-1} or as motivated by the Choi state in \cref{app:choi}, as
\begin{align}
E_{\mathrm{op}}(X) \equiv -\text{log}\left( \frac{\operatorname{Tr}\left[ \mathbb{S}_{AA'} X^{\otimes 2} \mathbb{S}_{AA'} X^{\dagger \otimes 2} \right]}{\left\Vert X \right\Vert_{2}^{4}} \right),
\label{eq:2-renyi-ent}
\end{align}
where the log is base two. This reduces to the usual definition related to \cref{eq:lin_herm_opent} if \(X\) is unitary. We choose this particular form over linear or log entropy because it is efficient in calculations while still scaling extrinsically for chaotic systems. 

A rich physical scenario which is described by an effective non-Hermitian Hamiltonian is that of a quantum system subject to continuous measurement and postselection \cite{gopalakrishnan_entanglement_2021}. Broadly speaking, if the underlying unitary dynamics is nonintegrable then a typical initial state, for weak measurement strength, will grow into a volume-law phase, while, for strong measurements, it will grow into an area-law phase; see Refs. \cite{PhysRevB.98.205136,PhysRevX.9.031009,PhysRevB.99.224307,PhysRevB.100.134306,PhysRevX.10.041020,PhysRevLett.125.030505,PhysRevX.11.011030} for a detailed discussion. It is important to develop physically motivated quantifiers for these \textit{measurement-induced} entanglement phase transitions and, as we will show now, the operator entanglement of the time evolution operator \textit{itself} undergoes a phase transition as the measurement strength is varied. This section discusses the transition broadly, while a more detailed analysis is provided in sections \ref{sec:LTA} and \ref{sec:spectral}.

\begin{figure*}[!ht]
\raggedright
\begin{subfigure}{.45\textwidth}
\includegraphics[width=1\linewidth]{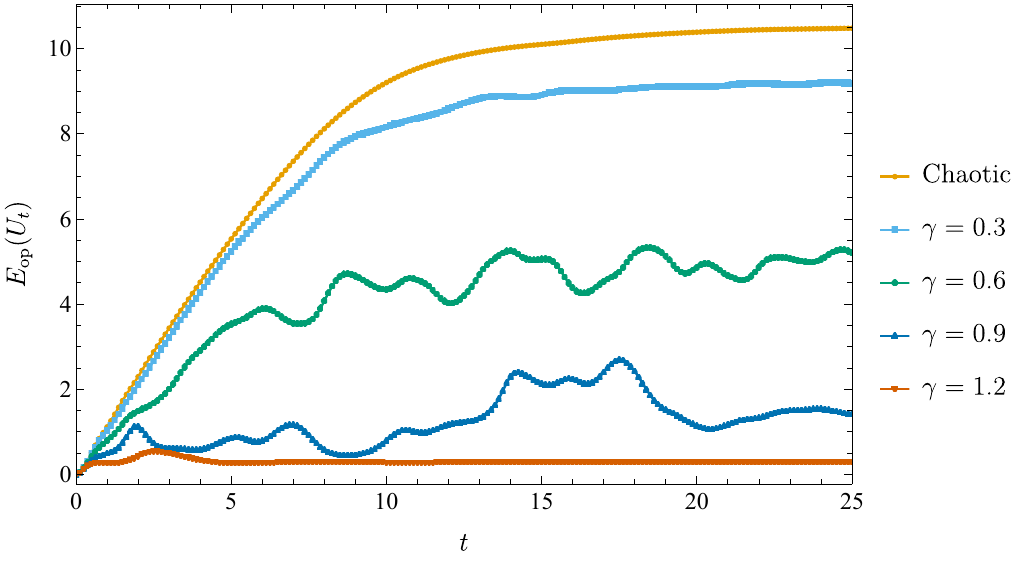}
\caption{}
\end{subfigure}\hspace{40pt}
\begin{subfigure}{.45\textwidth}
\includegraphics[width=1\linewidth]{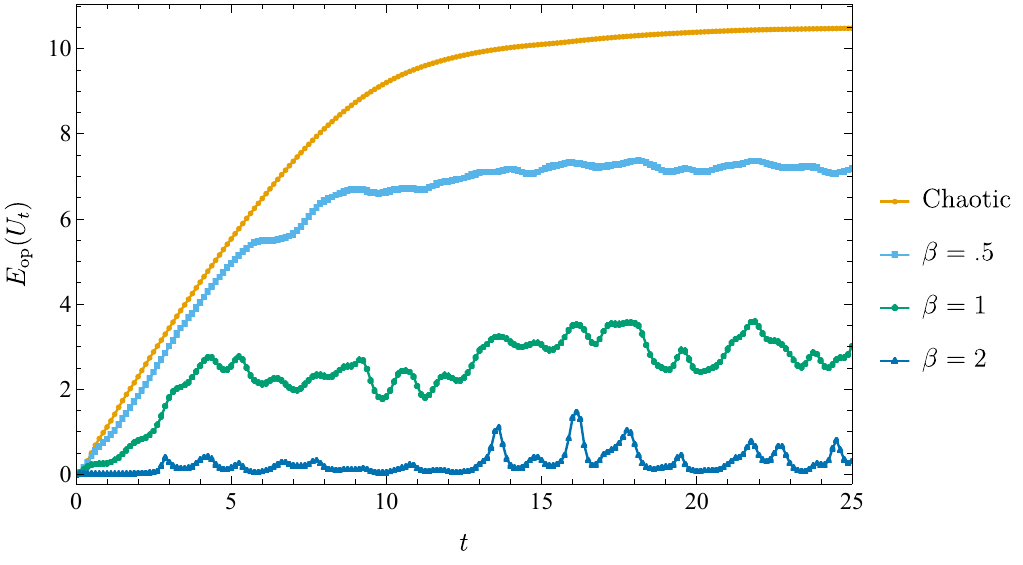}
\caption{}
\end{subfigure}
\par\bigskip
\begin{subfigure}{.45\textwidth}
\includegraphics[width=1\linewidth]{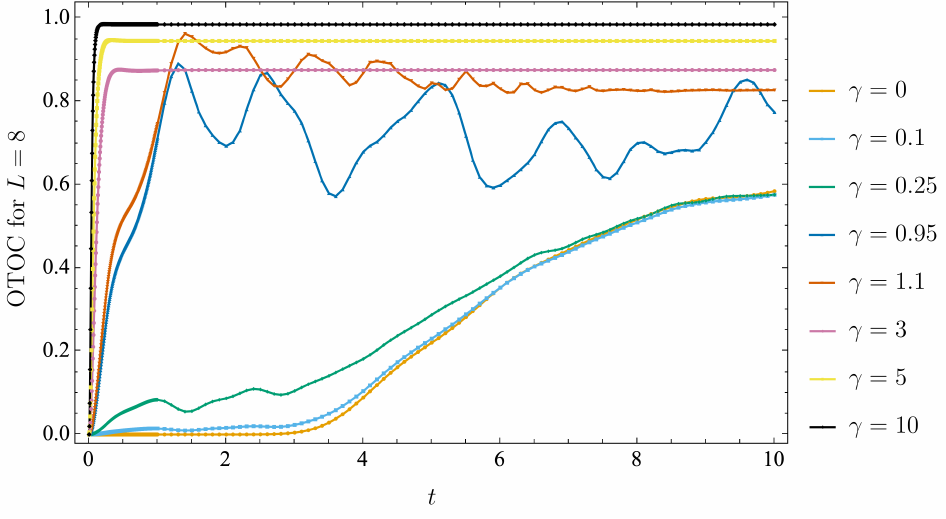}
\caption{}
\end{subfigure}\hspace{40pt}
\begin{subfigure}{.45\textwidth}
\includegraphics[width=1\linewidth]{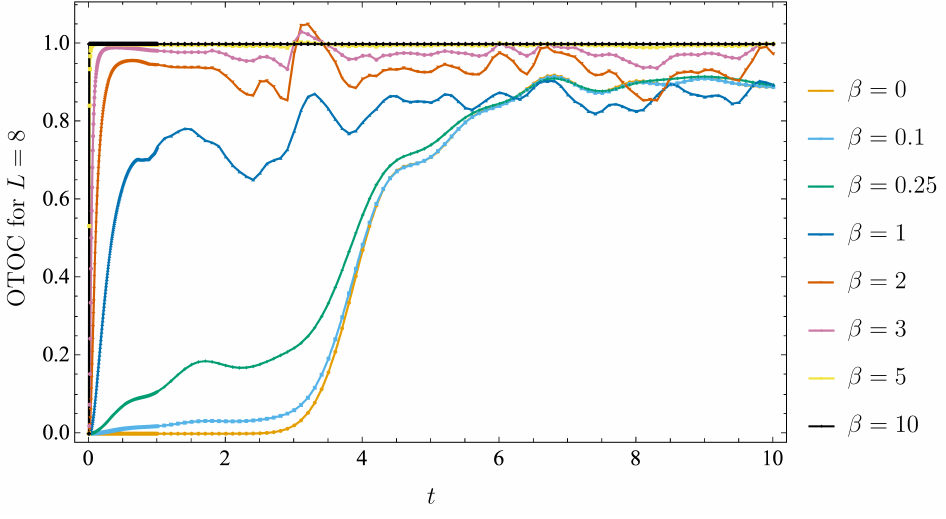}
\caption{}
\end{subfigure}
\caption{Operator entanglement as a function of time for the (a) measurement-induced and (b) isospectral non-Hermitian chaotic TFIM at $L=12$. The measurement-induced model displays three phases: the chaotic phase near the Hermitian limit, the integrable phase for $\gamma = 0.9$, and the purification phase for $\gamma = 1.2$. In contrast, the isospectral models appears to undergo a smooth transition from chaotic to integrable as $\beta$ is increased. Figures (c) and (d) correspond to the normalized OTOC (see \cref{eq:OTOC-namit}) for the two models at size \(L=8\) qubits. The operators \(V = \sigma_x = W\) are local operators at sites \(1,L\) (ends of the 1D chain) and the OTOC saturates polynomially fast to its long-time value at large non-Hermiticity \(\beta,\gamma\). The breakdown of LR bounds is also evident from figures (c), (d) since the normalized OTOC starts to grow at \(t \gtrapprox 0\) for larger values of \(\beta,\gamma\).}
\label{fig:TFIM-chaos}
\end{figure*}

We first investigate this transition numerically by plotting the growth of $E_\mathrm{op}(U_t)$ for the measurement-induced and isospectral extensions of the chaotic, integrable, and classical TFIMs, where $U_t = e^{-i H t}$ is the generally non-unitary time evolution operator. 

Plots of $E_\mathrm{op}(U_t)$ for the measurement-induced and isospectral non-Hermitian extensions of the chaotic TFIM are given in \cref{fig:TFIM-chaos}. In the chaotic phase $E_\mathrm{op}(U_t)$ quickly saturates to its long-time average value, which is $\overline{E_\mathrm{op}(U_t)} \approx L-1.6$ for the Hermitian chaotic TFIM with system size $L$ \cite{anand_brotocs_2021_published}. In each case increasing nonhermiticity parameter transitions the operator entanglement from quick saturation to suppressed oscillations, typical of integrable systems. The measurement-induced model alone further undergoes a purification phase transition at the exceptional point at $\gamma \gtrsim 1$, where the spectrum becomes complex. In the purification phase nontrivial behavior occurs only for small $t \lesssim 5$, as $U_t$ suppresses all but the one-dimensional long-time eigenspace exponentially in time.

\begin{figure*}[!ht]
\raggedright
\begin{subfigure}{.45\textwidth}
\includegraphics[width=1\linewidth]{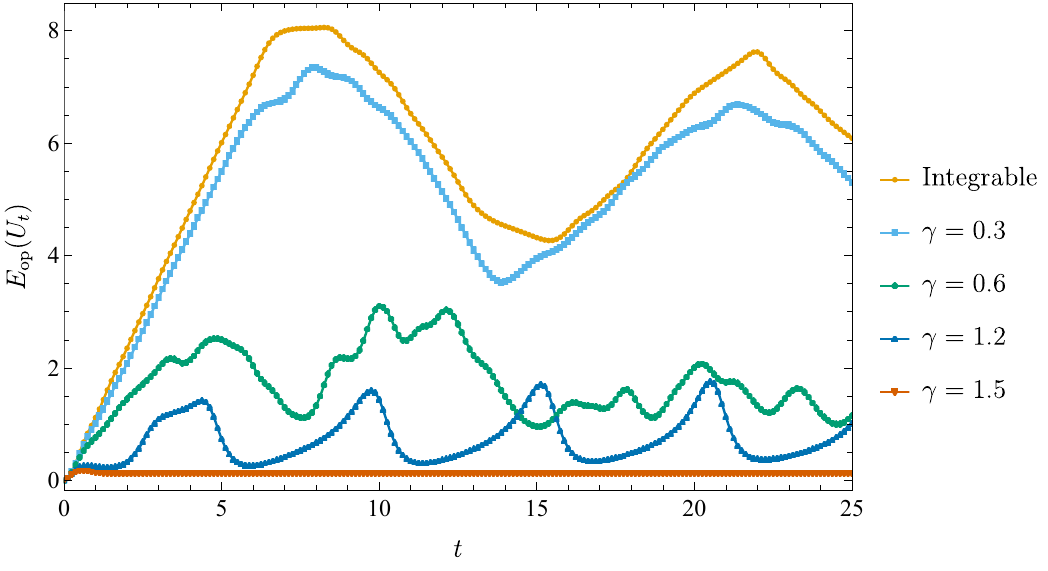}
\caption{}
\end{subfigure}\hspace{40pt}
\begin{subfigure}{.45\textwidth}
\includegraphics[width=1\linewidth]{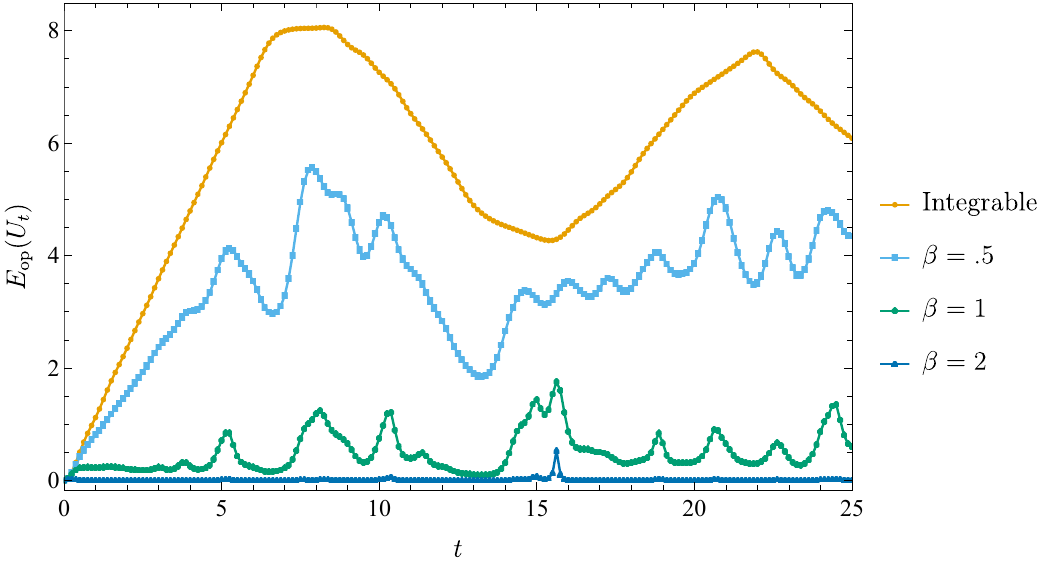}
\caption{}
\end{subfigure}
\caption{Operator entanglement as a function of time for the measurement-induced (a) and isospectral (b) non-Hermitian extensions of the integrable TFIM for $L=12$. The line corresponding to $\gamma=0.9$ overlaps that of $\gamma=1.2$ and is omitted for visibility. Unlike in the chaotic case, $\gamma=1.2$ generates highly periodic operator entanglement, owing to the persisting spectral degeneracy at this point. The isospectral model acts similarly to the chaotic case as $\beta$ is increased.}
\label{fig:TFIM-int}
\end{figure*}

Similar effects of nonhermiticity, including the purification transition, are seen for extensions of the integrable TFIM in \cref{fig:TFIM-int}. Unlike in the chaotic case, the measurement-induced extension has nontrivial behavior at $\gamma=1.2$, even though the spectrum is complex. As we will see, this occurs because the additional symmetry in the Hamiltonian induces a degenerate long-time eigenspace after the first exceptional point, allowing for more complex behavior. At $\gamma=1.5$ a second set of exceptional points has broken this symmetry, creating a nondegenerate long-time eigenspace and trivial long-time behavior. 

\begin{figure}[!ht]
\raggedright
\includegraphics[width=1\linewidth]{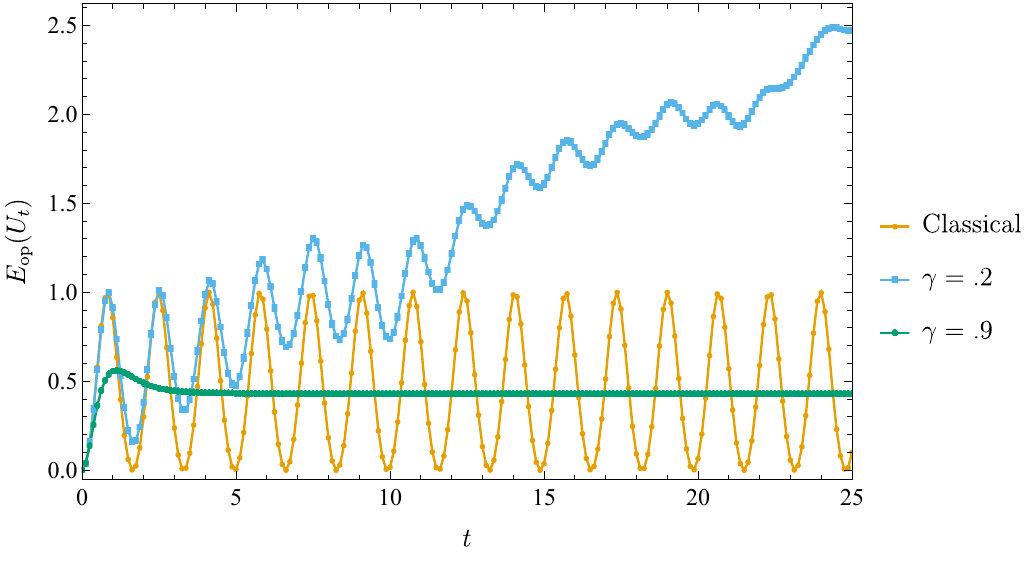}
\caption{Operator Entanglement of the measurement-induced extension of the classical TFIM with $h=0$ and $L=12$. As $\gamma$ is increased we see two phase transitions, the first when nonhermiticity makes the model become quantum and causes operator entanglement growth, and the second when the models shifts into the purification phase. In the latter, one still sees brief operator entanglement growth before the exponential decay of transient eigenspaces dominates.}
\label{fig:TFIM-cla}
\end{figure}

$E_\mathrm{op}(U_t)$ for the non-Hermitian extensions of the classical TFIM are plotted in \cref{fig:TFIM-cla}, where we see the same transition from integrable to purification phase as before, but at lower $\gamma$. Like the chaotic and integrable models the classical model has a degeneracy point at $\gamma = g = 0$, but here it is most clear that this is not an exceptional point. In fact, for $\gamma = 0.2$ we see aperiodic $E_\mathrm{op}$ growth, indicating a third, properly quantum, phase. 

Further determination of the phases comes from examining operator entanglement saturation values, which we investigate using system-size scaling and nonhermiticity dependence of the $E_\mathrm{op}(U_t)$ LTA, which closely measures this saturation value.

\subsection{Quantum quenches}
\label{subsec:quench}
The thermalization of closed quantum systems is a long-standing problem that has initiated a number of analytical, numerical, and experimental investigations \cite{Mitra_2018,gogolinEquilibrationThermalisationEmergence2016}. A number of these studies have focussed on the phenomena of quantum quenches, where starting from an initial eigenstate of a (pre-quench) Hamiltonian, one suddenly applies a (post-quench) Hamiltonian. The noncommutativity of the pre- and post-quench Hamiltonian ensures that the initial eigenstate has nontrivial time evolution. The experimental accessibility of quenched dynamics in various quantum simulators has fueled their study, in particular focussing on the growth of entanglement and correlations. While the exact dependence of post-quench entanglement growth rate is a function of the initial state, the integrability (or lack thereof) of the Hamiltonian, and so on, for local Hermitian Hamiltonians with bounded interaction strength, the entanglement growth can be \textit{at most} linear in time \cite{alba_entanglement_2018}. A proof of this relies on the LR bound, see, e.g., Ref.~\cite{bravyi_lieb-robinson_2006}. The violation of this bound for local non-Hermitian systems invites the question of whether entanglement growth can be beyond linear for these systems, which we answer affirmatively. 

To study quantum quenches in non-Hermitian systems we will focus on the post-selected evolution, and therefore the quantum state is pure at all times, cf. \cref{eq:pure-evolution}. The initial state will be a Neel state on \(L\) qubits, i.e., of the form, \(| \Psi(0) \rangle = | 0101 \cdots 01 \rangle\). Then,
\begin{align}
| \Psi(t) \rangle = \frac{U_{t} | \Psi(0) \rangle}{\left\Vert U_{t} | \Psi(0) \rangle \right\Vert_{}^{} }
\end{align}
describes the effective postselected evolution. This ensures that the state is pure at all times and therefore we can study the entanglement entropy of \textit{non-Hermitian} quenches in a similar way to the standard approach. That is, consider a partition of the Hilbert space, \(A|B\) and define the reduced state \(\rho_{A}(t) := \operatorname{Tr}_{B}\left[ | \Psi(t) \rangle \langle  \Psi(t) |  \right]\) and the quenched entanglement entanglement entropy, \(S_{\mathrm{EE}}(t) = S(\rho_{A}(t))\), where \(S(\cdot)\) is the von Neumann entropy of the reduced density matrix.

In particular, our quench studies are focused on \textit{interacting} non-Hermitian Hamiltonians, as opposed to the mostly noninteracting case studied in previous works \cite{PhysRevLett.120.185301,PhysRevLett.128.146804,Biella2021manybodyquantumzeno,PhysRevX.4.041001,pi2021phase}. This in turn also means that the typical analytically tractable techniques, e.g., mapping to free fermions via Jordan-Wigner transformation are not directly applicable to our work \cite{mbeng_quantum_2020}. To study quantum quenches numerically, we utilize the powerful time-evolving block decimation technique \cite{vidalEfficientClassicalSimulation2003} in the ITensor package \cite{itensor}. We focus on a maximum bond dimension of \(D = 256\) for our numerics, which, in the chaotic case and \(\beta \lessapprox 1\) necessarily leads to a truncation of the time-evolving state. 

To understand the thermalization of local subsystems under the isospectral TFIM \cref{eq:tfim-isospectral}, we numerically study the entanglement entropy following a quench from a Neel state. We study (i) subsystems of increasing size $l$ for a fixed \(L\) and (ii) an equal bipartition for increasing \(L\). For the former one expects that small subsystems will thermalize while larger ones will still retain information about the initial conditions. The latter lets us quantify the rate of thermalization of a small subsystem as we approach the thermodynamic limit. For both, we focus on the integrable and chaotic limits of the isospectral TFIM, for various values of the measurement strength.

\begin{figure*}[!th]
\raggedright
\begin{subfigure}{.4\textwidth}
\includegraphics[width=1.3\linewidth]{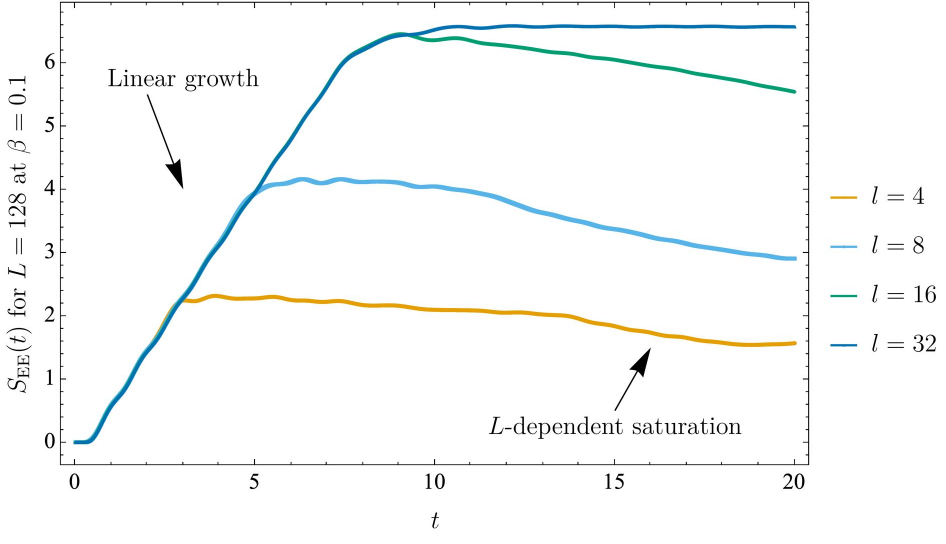}
\caption{}
\label{fig:sqivi-a}
\end{subfigure}\hspace{65pt}
\begin{subfigure}{.4\textwidth}
\includegraphics[width=1.3\linewidth]{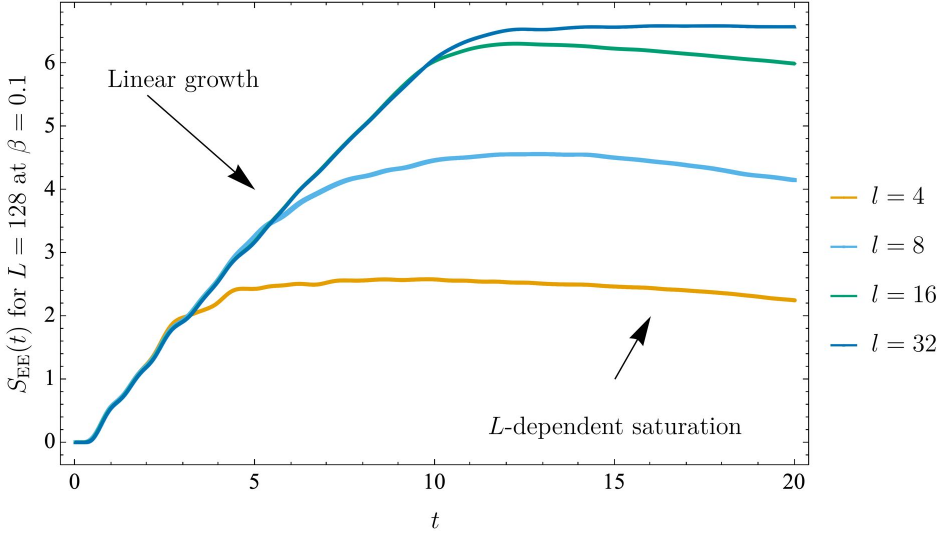}
\caption{}
\label{fig:sqivi-b}
\end{subfigure}%
\caption{Growth of entanglement entropy following a quantum quench from the Neel state for \(L=128\) and subsystem size $l=\{4,8,16,32\}$. The MPS simulations are for the isospectral TFIM at $\beta = 0.1$ for (a) the integrable model and (b) the chaotic model. At small measurement strength, we notice that both models seem to thermalize equally well for subsystem sizes up to $L/4$, with the equilibration value of the integrable quench being lower than the chaotic regime.}
\label{fig:shorttime-quench-subsystem-isospectral}
\end{figure*}

\begin{figure}[!ht]
\raggedright
    \includegraphics[width=1\linewidth]{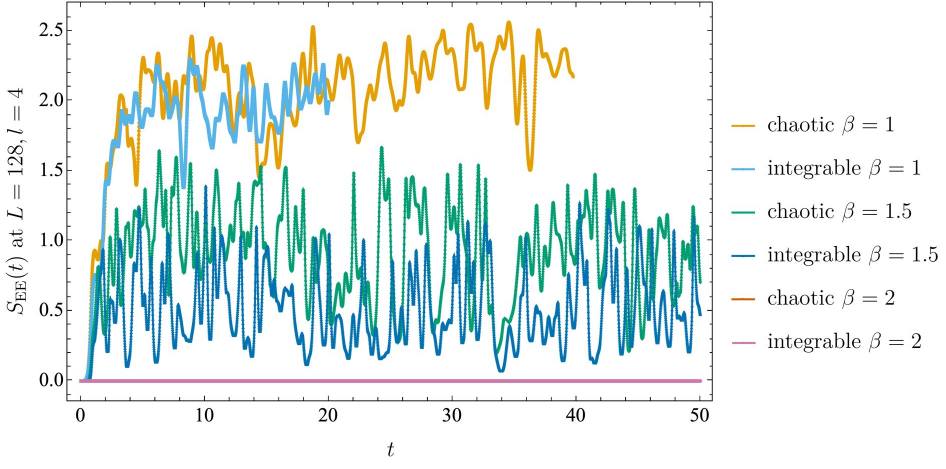}
    \caption{Growth of entanglement entropy following a quantum quench from the Neel state for \(L=128\) and a subsystem of size $l=4$. The MPS simulations are for the isospectral TFIM at $\beta = \{1,1.5,2\}$ for both the integrable and chaotic regimes. 
    Interestingly, we notice that the thermalization properties of both the chaotic and the integrable model vanish even at moderate non-Hermiticity strength $\beta \approx 1$, while both completely suppress entanglement at $\beta \gtrapprox 1$.}
\label{fig:quench-thermalization-large-beta}
\end{figure}

\begin{figure*}[!th]
\raggedright
\begin{subfigure}{.4\textwidth}
\includegraphics[width=1.3\linewidth]{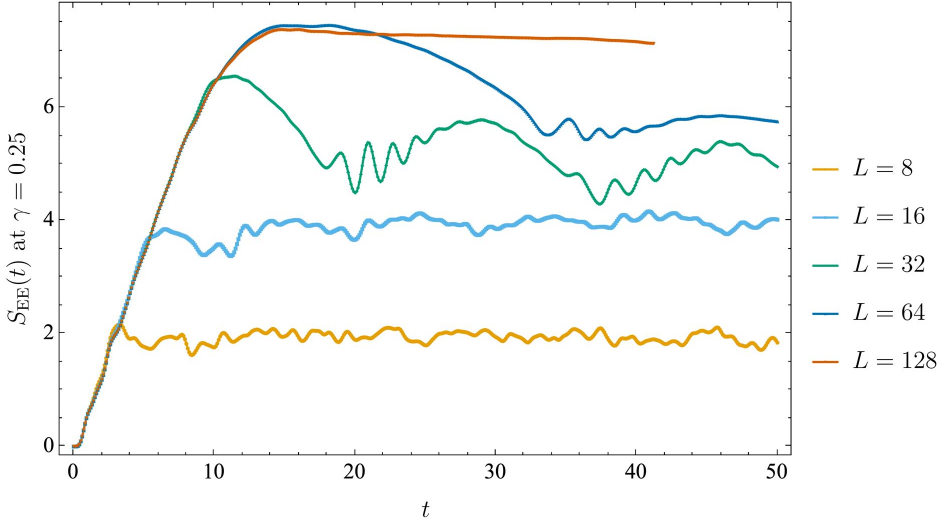}
\caption{}
\end{subfigure}\hspace{65pt}
\begin{subfigure}{.4\textwidth}
\includegraphics[width=1.3\linewidth]{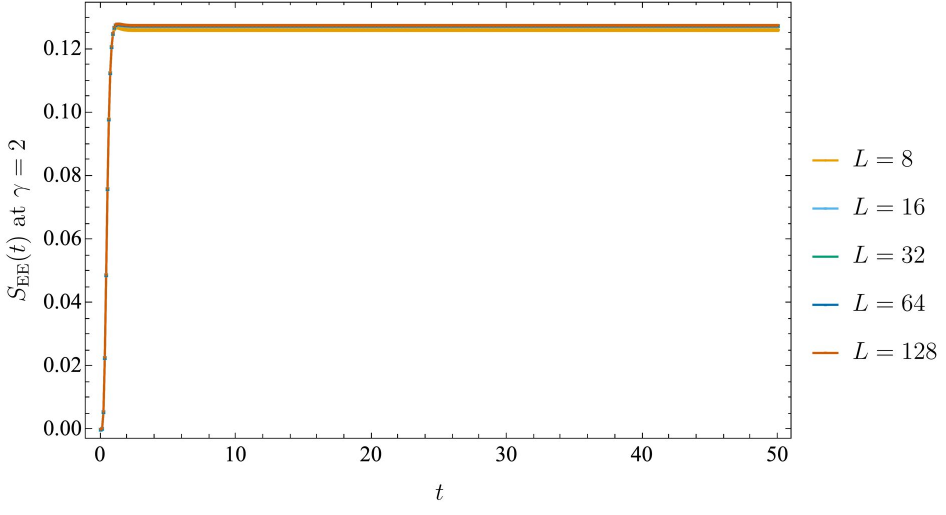}
\caption{}
\end{subfigure}%
\caption{Growth of entanglement entropy following a quantum quench from the Neel state for \(L=\{8,16,32,64,128\}\) and $L/2:L/2$ bipartition. The MPS simulations are for the measurement-induced TFIM in the (a) volume-law phase at \(\gamma = 0.25\) and (b) area-law purification phase at \(\gamma = 2\).}
\label{fig:shorttime-quench-imaginary}
\end{figure*}

In \cref{fig:shorttime-quench-subsystem-isospectral}, we note that both the integrable and chaotic phases at small non-Hermiticity, $\beta = 0.1$ display a thermalization of the subsystem, up to size $l=L/4$. In contrast, we notice that at $\beta \gtrapprox 1$, cf. \cref{fig:quench-thermalization-large-beta} both chaotic and integrable models seem to fail thermalization of even small subsystems, in this case of size $l=4$. In \cref{fig:shorttime-quench-imaginary} we see that while the measurement-induced model has $L$-dependant entanglement saturation at small measurement rate $\gamma=0.25$, the saturation value is $L$-independent in the purification phase with $\gamma=2$.

\section{Long-time averages}
\label{sec:LTA}

In Hermitian systems, the long-time average (LTA) of the operator entanglement can distinguish integrable and chaotic models, e.g. via its system-size scaling, making it an interesting time-invariant quantity to consider for non-Hermitian systems \cite{styliaris_information_2021,anand_brotocs_2021_published}.

\subsection{Analytic approximation}

The operator entanglement LTA can be taken analytically in Hermitian systems, given the so-called non-resonance condition (NRC) \cite{styliaris_information_2021}. The NRC requires that differences between eigenvalues be unique, and can be thought of as a stronger extension of nondegeneracy. In non-Hermitian systems however, time dependence of the normalization term in time-evolved quantities prevents exact analytic calculation, but in certain cases the LTA can be analytically approximated sufficiently well to study the necessary physics. 

Here, the specific quantity we are interested in is the LTA of the 2-Renyi entropy (\cref{eq:2-renyi-ent}) of the time evolution operator, $E_{\mathrm{op}}(U_t)$. 

First, let us assume $H$ is diagonalizable and consider the form of $U_t$ given in \cref{eq:diagonal}. Notice that any constant imaginary shift in $\lambda_i$ will affect both the numerator and denominator of $E_\mathrm{op}(U_t)$ equivalently, and so will cancel out in the overall fraction. Without loss of generality we can then assume all $\text{Im}[\lambda_i] \leq 0$ with $\text{max}(\text{Im}[\lambda_i])=0$. As the LTA is invariant under short time behavior, we can take the limit $t \rightarrow \infty$ before averaging. All eigenspaces with $\text{Im}[\lambda_i] < 0$ are exponentially suppressed and do not contribute in this limit, and we are left with only terms that depend on the long-time eigenspace $\mathcal{H}_L$.

The approximation we must make is $\overline{\text{log}(f)} \approx \text{log}(\hspace{1mm}\overline{f }\hspace{1mm})$, which comes from typicality of Haar-distributed unitaries and is valid for chaotic systems. For non-chaotic systems it is qualitatively correct, and sufficiently accurate to distinguish phases of non-Hermitian systems. As $\text{log}(f/g)=\text{log}(f)-\text{log}(g)$, this approximation allows the numerator and denominator to be averaged independently within the logarithm.

Additionally, we must assume nonresonance in $\mathcal{H}_L$ - a weaker condition than full NRC, as it only applies to eigenvalues with maximal imaginary component. Within $\mathcal{H}_L$ all eigenvalues have only real differences, so long-time NRC allows an average to be taken using techniques from \cite{styliaris_information_2021}. The approximate LTA of $E_\mathrm{op}(U_t)$ is then derived in Appendix~\ref{app:LTA-derivation}, and found to be 

\begin{widetext}
\begin{align}
\overline{E_{\mathrm{op}}(U_t)} \approx -\log \frac{\operatorname{Tr}\left[ R_A L_A \right] + \operatorname{Tr}\left[ R_B L_B \right] - \operatorname{Tr}\left[ \mathrm{diag}(R_A) \mathrm{diag}(L_A) \right]}{[\text{Tr}[\eta_{_L}]]^2 + \text{Tr}[R L] - \text{Tr}[L]}
\label{eq:LTA-Analytic}
\end{align}
\end{widetext}

where we define 

\begin{align}
\label{eq:reduced-Gram}
\begin{split}
    (R_X)_{i,j} &\equiv \langle \rho^X_i, \rho^X_j \rangle\\
    (L_X)_{i,j} &\equiv \langle \sigma^X_i, \sigma^X_j \rangle
    \end{split}
\end{align}

for $\vert i\rangle,\vert j\rangle\in\mathcal{H}_L$ as the long-time right and left reduced Gram matrices, which remain symmetric in this context, and 
\begin{align}
\label{eq:reduced-eigenstates}
\rho_k^X \equiv \text{Tr}_{\bar X}\left( \vert  r_{k} \rangle \langle  r_{k} \vert \right),\ \sigma_{j}^X \equiv \text{Tr}_{\bar X} \left(\vert l_{j} \rangle \langle  l_{j} \vert \right)
\end{align}
as the reduced left and right Hamiltonian eigenstates. In the denominator, $\eta_{_L} = \sum_{i\in\mathcal{H}_L} \vert l_i\rangle\langle l_i\vert$ is the restriction to $\mathcal{H}_L$ of the Hilbert space metric $\eta$ satisfying $H^\dag \eta = \eta H$. $L$ and $R$ are the equivalent unreduced Gram matrices where no partial trace is taken. In the Hermitian limit \cref{eq:LTA-Analytic} reduces the the form given in Ref.~\cite{styliaris_information_2021}.

Of the models here considered, the chaotic TFIM obeys NRC outside of its degenerate and exceptional points. The integrable TFIM never obeys true NRC, and only obeys long-time NRC when $\mathcal{H}_L$ is one-dimensional, e.g. at large $\gamma$. The classical model is degenerate in the Hermitian limit, but obeys NRC for any $\gamma > 0$.

Error in this formula arises from nonorthogonality of Hamiltonian eigenstates, which is maximal around the exceptional points where eigenstates become parallel, and from NRC violations, which are present at both degenerate and exceptional points. However, as shown in Appendix~\ref{app:LTA-analytic-plots} the formula captures the correct system-size scaling and qualitative dependence on nonhermiticity parameter of the entanglement LTA, as well as the correct qualitative behavior around the degeneracy and exceptional points. Thus, it can still be useful for detecting measurement-induced phase transitions and scaling behavior of non-Hermitian systems.

\subsection{Finite-size scaling}

We now investigate the usefulness of the operator entanglement LTA applied to physical models and random matrix ensembles. This was previously studied for Hermitian models in Ref.~\cite{anand_brotocs_2021_published}, where it was found that $\overline{E_\mathrm{op}(U_t)} \approx \mu L - \text{log}(\alpha)$, where $\mu\approx 1$ for the chaotic TFIM and GUE ensemble, $\mu\approx 0.5$ for the integrable TFIM and NRC Product State (``NRCPS'') ensemble, and $\alpha$ is a model-dependant constant. The NRCPS ensemble is described in Ref.~\cite{anand_brotocs_2021_published}, and is an example of the most chaotic an integrable Hamiltonian can be.

\begin{figure}[!th]
\raggedright
    \includegraphics[width=1\linewidth]{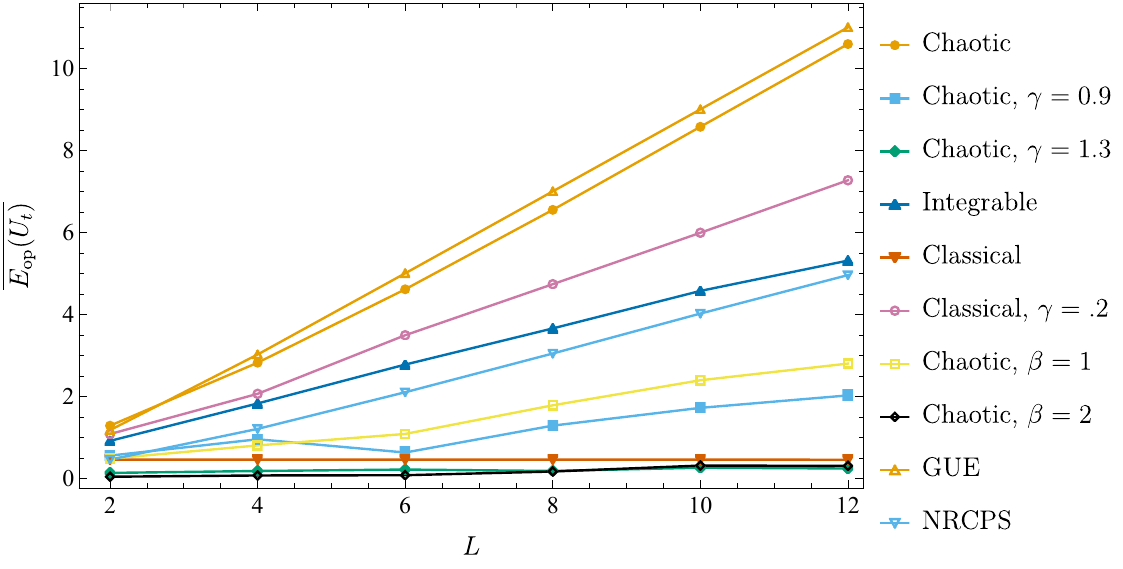}
    \caption{Comparison of system size scaling of the $E_\mathrm{op}(U_t)$ LTA for a range of models. The Gnibre ensemble is omitted, as its behavior is identical to that of the GUE. }
    \label{fig:scaling_comparison}
\end{figure}

Our numerical results are plotted in \cref{fig:scaling_comparison}, where we see three general regions of operator entanglement growth corresponding to the volume law chaotic models, suppressed volume law integrable models, and area law integrable models. The scaling of Hermitian models here mirrors the findings of Ref.~\cite{anand_brotocs_2021_published}\footnote{Though Ref.~\cite{anand_brotocs_2021_published} describes $\text{log}(\overline{1-E_\text{op}^\text{lin}})$ and we study $\overline{E_\text{op}}=\overline{\text{log}(1-E_\text{op}^\text{lin})}$, as mentioned earlier this is equivalent for sufficiently chaotic systems, as numerically confirmed here.}, and includes both volume-law models, and the chaotic TFIM and GUE. The second general region of growth corresponding to suppressed volume law integrable growth includes the NRCPS ensemble, the classical TFIM with $\gamma=0.2$, the chaotic TFIM with $\gamma=0.9$, and the two isospectral chaotic TFIM models, which all demonstrate mostly linear growth in $L$ but with reduced coefficients relative to the volume-law models. The isospectral model with $\beta=2$ fits into this region of growth because one can see for large $L$ it does faintly grow, unlike the classical TFIM and chaotic TFIM with $\gamma=1.3$. These last two models exhibit area law, in this case constant, growth and constitute the third region. Unlike the Hermitian case it is difficult to find a universal linear fit as entanglement LTA scaling can be nonlinear, and even non-monotonic, in $L$.

\begin{figure}[!th]
\raggedright
    \includegraphics[width=1\linewidth]{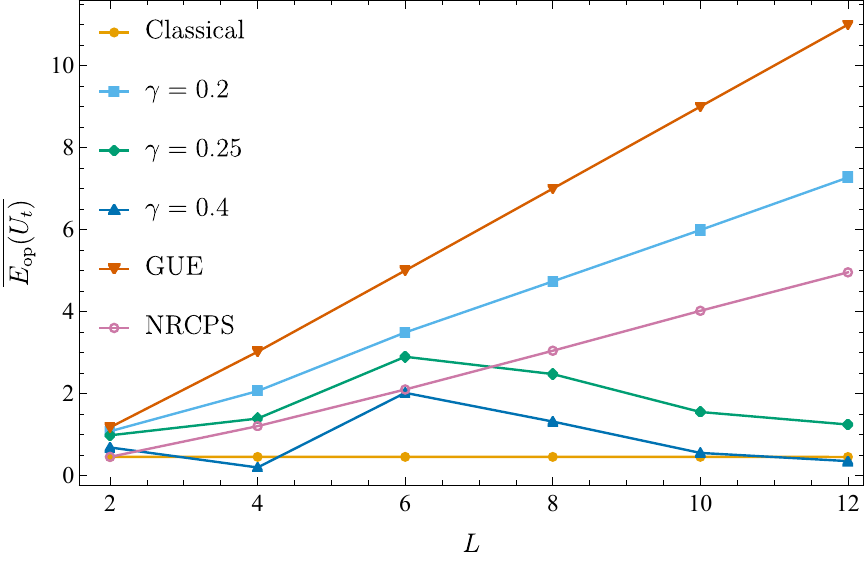}
    \caption{System size scaling of $E_\mathrm{op}(U_t)$ for the measurement-induced classical TFIM with $J=.95$, $g=0$, and $h=0.5$, with GUE and NRCPS averages included for comparison.}
    \label{fig:scaling_integrable}
\end{figure}

A closeup of the scaling behavior of the classical TFIM is plotted in \cref{fig:scaling_integrable}. For intermediate $\gamma=0.2$, just before the first exceptional point, the classical model shows greater scaling in system size than the NRCPS average, indicating a potential lack of integrability. However, immediately past the exceptional point at $\gamma\approx 0.25$ there is a lack of monotonic growth in $L$ as the system enters the purification phase.

\subsection{Dependence on non-Hermiticity parameter}

\begin{figure}[!th]
\raggedright
\includegraphics[width=1\linewidth]{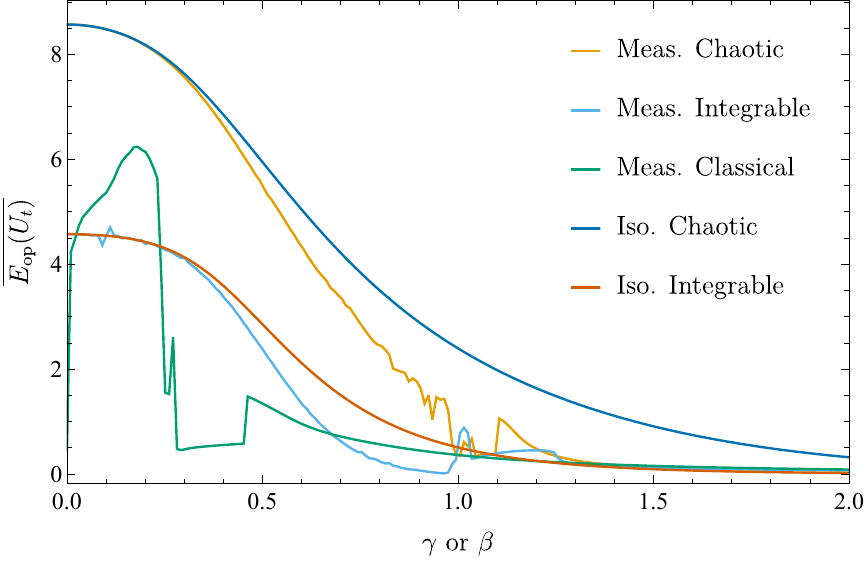}
    \caption{$E_{\mathrm{op}}$ LTA of various models as a function of nonhermiticity parameter ($\gamma$ or $\beta$) at $L=10$. Note that $\gamma$ and $\beta$ act differently as parameters and the two models should be compared only qualitatively. The dip in the classical model at $\gamma=0.2$ and other non-smoothness in the measurement-induced models past their exceptional points are likely due to eigenvalues becoming complex at different values of $\gamma$, causing changes in long-time eigenspace and discontinuities in saturation value.}
\label{fig:param_scaling}
\end{figure}

We use the operator entanglement LTA to study how $U_t$ changes as a function of nonhermiticity parameter, with numerical results plotted in \cref{fig:param_scaling}. Prior to the degenerate point, all nonclassicl models have similar $E_\mathrm{op}$ decay coming from increasing nonorthonormality of Hamiltonian eigenvectors, which are volume-law entangled in the Hermitian limit in the chaotic case. For the isospectral TFIMs, this is the only behavior that occurs, and the overall effect of nonhermiticity is monotonic decay of operator entanglement. On the other hand, the measurement-induced TFIM has non-monotonic scaling in $\gamma$ regardless of whether the Hermitian limit is chaotic, integrable, or classical. In each case, this occurs because the nonhermiticity breaks degeneracies present at the degenerate point ($\gamma=1$ in the chaotic and integrable models, $\gamma=0$ in the classical model), increasing operator entanglement immediately after this point. This growth continues until an exceptional point is reached, after which it is in competition with exponential suppression of subspaces from imaginary components of eigenvalues, which eventually dominates. This leads to non-monotonic growth and decay of entanglement after the exceptional point. 

In the integrable model alone, the degeneracy point is also a degenerate exceptional point, after which multiple eigenvalues take the same imaginary component (see \cref{fig:spectrum-int}). This creates a degenerate $\mathcal{H}_L$ and increased $\overline{E_\mathrm{op}(U_t)
}$ despite being in the purification phase. At the second exceptional point this degeneracy is broken, making $\mathcal{H}_L$ one-dimensional like in the other measurement-induced models.   

\section{Spectral analysis}
\label{sec:spectral}

\begin{figure*}[!ht]
\raggedright
\begin{subfigure}{.45\textwidth}
\includegraphics[width=1\linewidth]{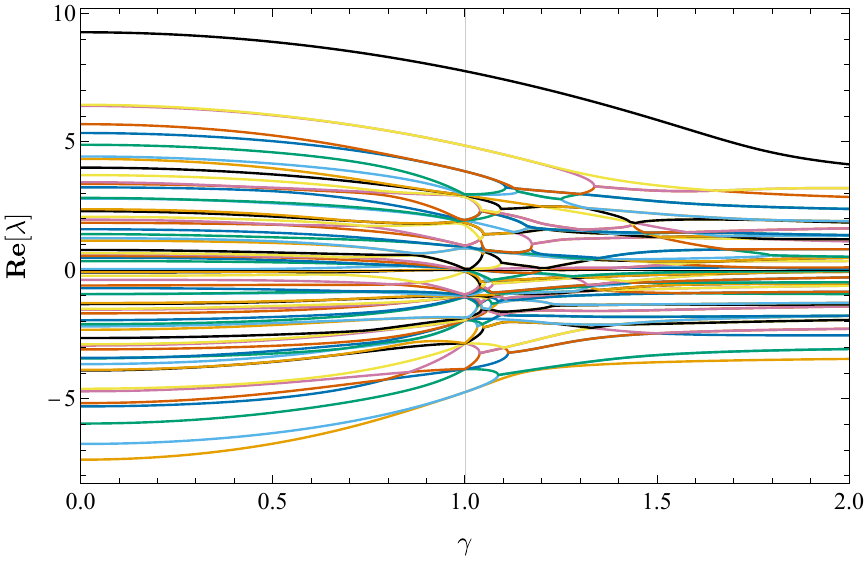}
\caption{Real part}
\end{subfigure}\hspace{20pt}
\begin{subfigure}{.45\textwidth}
\includegraphics[width=1\linewidth]{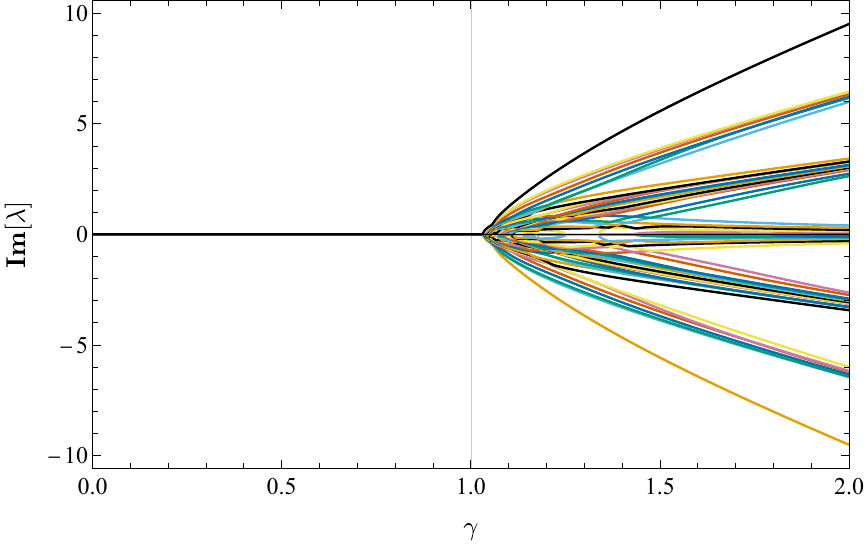}
\caption{Imaginary part}
\end{subfigure}
\caption{Spectrum of the chaotic TFIM with $L=6$ as $\gamma$ is increased. Here the degenerate point occurs at $\gamma=1$, which is marked with a gridline for clarity. Eigenvalues begin to go through exceptional points shortly thereafter. For large $\gamma$, $H_M$ is dominated by the $\gamma \sum_{i=1}^n \sigma_i^y$ term, which is why the eigenvalues converge towards $n+1$ distinct, linearly growing in $\gamma$, imaginary components, with static real parts.}
\label{fig:spectrum-chaotic}
\end{figure*}

\begin{figure*}[!th]
\raggedright
\begin{subfigure}{.45\textwidth}
\includegraphics[width=1\linewidth]{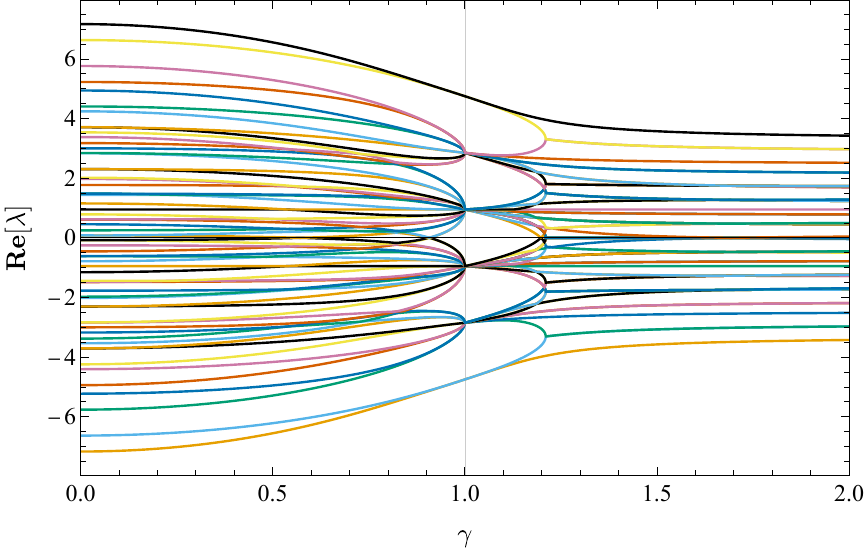}
\caption{Real part}
\end{subfigure}\hspace{20pt}
\begin{subfigure}{.45\textwidth}
\includegraphics[width=1\linewidth]{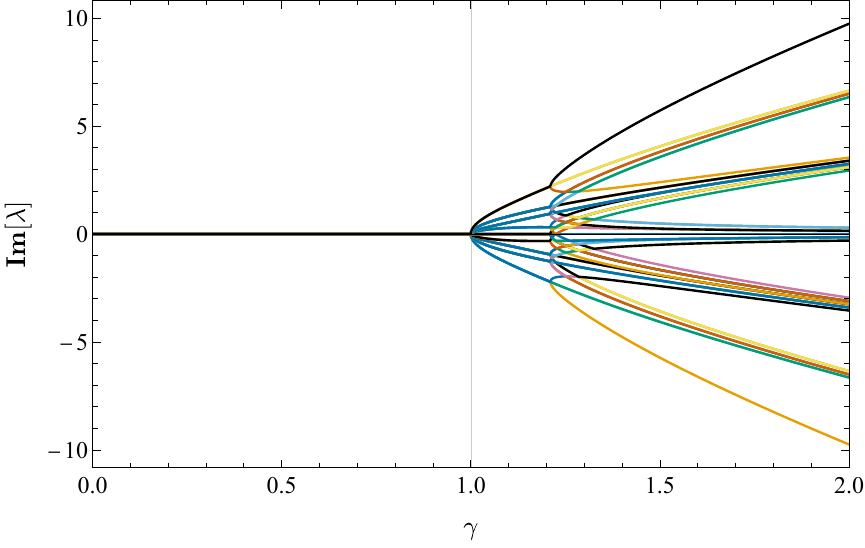}
\caption{Imaginary part}
\end{subfigure}
\caption{Spectrum of the integrable TFIM with $L=6$ as $\gamma$ is increased. The degeneracy point occurs the same as in the chaotic case, but now coincides with the first exceptional point. There is additional degeneracy in eigenvalues that exists in the Hermitian limit and remains unbroken until the second exceptional point at $\gamma\approx 1.2$. Unlike both the chaotic and classical models, all exceptional points occur at two discrete values of $\gamma$}
\label{fig:spectrum-int}
\end{figure*}

\begin{figure*}[!ht]
\raggedright
\begin{subfigure}{.45\textwidth}
\includegraphics[width=1\linewidth]{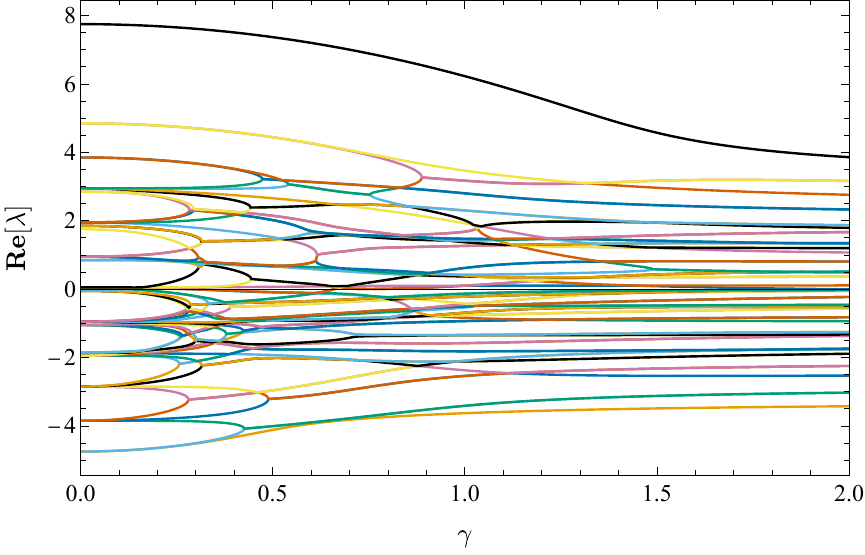}
\caption{Real part}
\end{subfigure}\hspace{20pt}
\begin{subfigure}{.45\textwidth}
\includegraphics[width=1\linewidth]{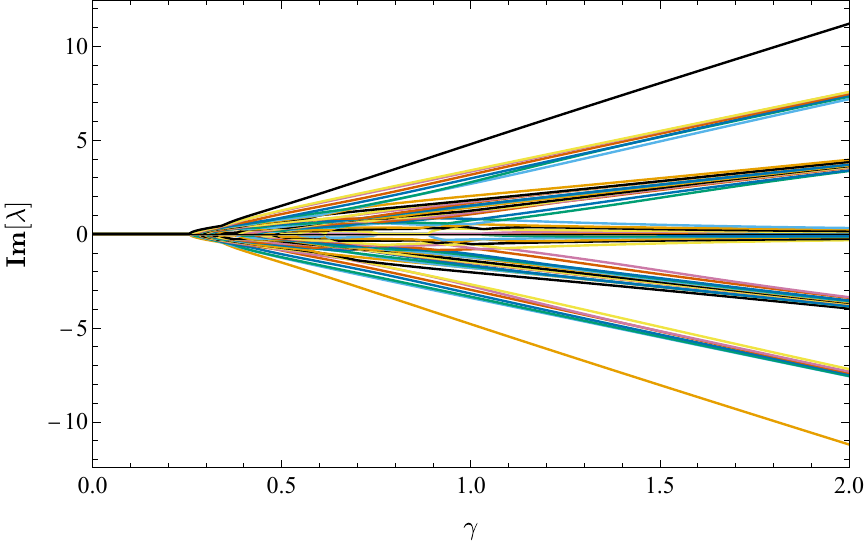}
\caption{Imaginary part}
\end{subfigure}
\caption{Spectrum of the classical TFIM with $L=6$ as $\gamma$ is increased. Here the degeneracy point occurs in the Hermitian limit, and the gap between that and the first exceptional point is much larger than in the chaotic case.}
\label{fig:spectrum-cla}
\end{figure*}

The measurement-induced phase transition in the measurement-induced chaotic TFIM was studied in Ref.~\cite{gopalakrishnan_entanglement_2021} in terms of the level spacing ratio, and standard spectral measure of chaoticity. This quantity was found to be hard to generalize to the case of complex spectra. However, direct inspection of Hamiltonian spectrum for simple cases is sufficient to understand the non-monotonicity in the measurement-induced TFIM seen in \cref{fig:param_scaling}. To this end, the spectra of the measurement-induced chaotic, integrable, and classical TFIM are plotted in Figs.~\ref{fig:spectrum-chaotic}, \ref{fig:spectrum-int}, and \ref{fig:spectrum-cla}, respectively.

From \cref{fig:spectrum-chaotic} we see that for the measurement-induced chaotic TFIM, as $\gamma$ approaches the degenerate point at $\gamma=g$ from the left, most eigenvalues converge towards a few unique values, reducing the effective eigenvalue range and thus chaoticity of the system. Past this degenerate point, the eigenvalues split again, but begin to go through exceptional points and pick up imaginary components, causing the purification phase. These imaginary components are seen to be nondegenerate, leading to a one-dimensional long-time eigenspace. The competing effects of eigenvalue splitting and eigenspace decay lead to the non-monotonic behavior of the $E_\mathrm{op}$ average after the degeneracy point seen in \cref{fig:param_scaling}.

In \cref{fig:spectrum-int} we see that the degeneracy of the integrable TFIM remains unbroken by nonhermiticity. Additionally, unlike the chaotic model, all exceptional points occur at one of two distinct $\gamma$ values, the first of which coincides with the degenerate point at $\gamma=g$. Between the two exceptional points the system is nominally in the purification phase, but eigenvalues with degenerate maximal imaginary components causes a multi-dimensional long-time eigenspace. This leads to unique plateu behavior in the $\overline{E_\mathrm{op}(U_t)}$ for $1\leq\gamma\lesssim 1.2$, as seen in \cref{fig:param_scaling}. This is the only model studied here which ever has a multi-dimensional long-time eigenspace other than the full Hilbert space. 

\cref{fig:spectrum-cla} shows the spectrum of the classical TFIM, which appears as a special case of the chaotic TFIM with degeneracy point at $\gamma = g = 0$, the Hermitian limit. In contrast to the integrable model, $\gamma > 0$ acts as a perturbation, breaking degeneracies of the original $H_\text{TFIM}$ and making the spectrum no longer area-law. This leads to the sharp initial growth of $\overline{E_\mathrm{op}(U_t)}$ in \cref{fig:param_scaling}.  

\section{Concluding remarks}

Non-Hermitian Hamiltonians provide a fundamentally new avenue to explore well-established ideas in many-body physics. While the constraints induced from unitarity and locality have been understood quite rigorously, hybrid quantum circuits incorporating measurements require a revision of our intuition, especially due to the breakdown of LR bounds. In this paper we focused on aspects of information scrambling and quantum chaos of local, non-Hermitian variants of paradigmatic spin-chain models. The breakdown of LR bounds makes traditional OTOCs unwieldy as we exemplify with Fig. \ref{fig:lieb-robinson-9qubits-isospectral}. In contrast, we show that the operator entanglement -- which in the unitary case is closely related to OTOCs -- is still able to distinguish the scrambling properties of these Hamiltonians. Additionally, the long-time average of operator entanglement is able to distinguish the chaotic, integrable, and purification phases.

Moreover, quantum quenches allow us to understand the thermalization properties of the local subsystems and we note that, even at small measurement rates, the subsystems do not seem to thermalize. Operator entanglement is closely tied to simulability for Hermitian quantum circuits \cite{dubail_entanglement_2017}. The methods studied here prove a potential way to generalize this to hybrid quantum circuits, which may live in a separate computational complexity class from standard quantum circuits. This has potential use towards e.g. quantum supremacy experiments with non-Hermitian systems, where demonstrating the breakdown of simulability is required \cite{preskill_2012}. As a future direction, it would be worth exploring the classical and quantum computational complexity of simulating non-Hermitian Hamiltonians, in particular, following the imaginary time evolution protocol of Ref.~\cite{Motta_2019}.

\prlsection{Experimental prospects} The non-Hermitian Hamiltonians and their scrambling properties that are discussed in this paper can be readily explored in a variety of quantum simulators such as those based on ultracold atoms \cite{gross2017quantum}, trapped ions \cite{blatt2012quantum}, cavity QED \cite{schmidt2013circuit}, etc. This is owing to the natural spontaneous decay processes that occur in these systems, see the excellent discussion in Refs. \cite{PhysRevX.4.041001,PhysRevLett.113.250401}. To simulate the non-Hermitian evolution, \(e^{-iH_\text{I}t}\), we simply notice that \(e^{-iH_\text{I}t} = S(\beta) U_{t} S(\beta)^{-1}\) where \(U_{t}\) is the purely unitary dynamics. Since we are mostly interested in Ising-type Hamiltonians, these can be easily implemented in most current day experimental platforms. Moreover, to implement the nonunitary piece, \(S(\beta)\) we need to postselect on no quantum jumps under continuous monitoring. In particular, the ability to continuously monitor the cavity in 2D and 3D superconducting QPUs \cite{krantz2019quantum,alam2022quantum} would allow for an experimental verification of these non-Hermitian scrambling effects; although, this would probably require having multiple transmon qubits (or \textit{qudits}) coupled to the same cavity. Similarly, photonic systems provide a natural testbed for these experiments because of their ability to carefully tune gain and loss \cite{feng2017non}.

Furthermore, a rigorous approach to benchmarking quantum computing platforms is to quantify their ability to prepare computationally nontrivial states, for e.g., long-range entangled states. Generically, LR bounds prevent the preparation of such states by low-depth quantum circuits. However, in recent years, there has been a striking development in this area by the introduction of ``adaptive quantum circuits,'' unitary circuits interspersed with measurements, that violate the LR bound (this is also the class of dynamics that we focus on in this paper). Perhaps unsurprisingly, this allows for a low-depth preparation of long-range entangled states \cite{PRXQuantum.3.040337}. In fact, the approach in Ref. \cite{PRXQuantum.3.040337} is analogous to the ideas in fusion-based quantum computation \cite{bartolucci_fusion-based_2021}. Moreover, the introduction of measurements and postselection have inspired revisiting the existing classification of phases of matter \cite{PhysRevLett.127.220503}. We expect these ideas to provide further insights into benchmarking and characterization of quantum devices, both from a technological and foundational perspective.\\

\section{Acknowledgments}
N.A. would like to thank Arul Lakshminarayan and Himanshu Badhani for many insightful discussions. N.A. is a KBR employee working under the Prime Contract No. 80ARC020D0010 with the NASA Ames Research Center. The United States Government retains, and by accepting the article for publication, the publisher acknowledges that the United States Government retains, a nonexclusive, paid-up, irrevocable, worldwide license to publish or reproduce the published form of this work, or allow others to do so, for United States Government purposes.

The authors acknowledge the Center for Advanced Research Computing (CARC) at the University of Southern California for providing
computing resources that have contributed to the research results
reported within this publication. URL: \url{https://carc.usc.edu}. The
authors acknowledge partial support from the NSF award PHY-1819189. This research was (partially) sponsored by the Army Research Office and was accomplished under Grant Number W911NF-20-1-0075. The views and conclusions contained in this document are those of the authors and should not be interpreted as representing the official policies, either expressed or implied, of the Army Research Office or the U.S. Government. The U.S. Government is authorized to reproduce and distribute reprints for Government purposes notwithstanding any copyright notation herein.

J.M. is thankful for support from NASA Academic Mission Services, Contract No. NNA16BD14C. The NASA team's work (N.A., J.M., E.R.) was primarily funded by DARPA under IAA 8839 Annex 129, while the material on applications to benchmarking, simulation, and state preparation is based upon work supported by the U.S. Department of Energy, Office of Science, National Quantum Information Science Research Centers, Superconducting Quantum Materials and Systems Center (SQMS) under contract number DE-AC02-07CH11359 through NASA-DOE interagency agreement SAA2-403602.


\onecolumngrid
\widetext

\appendix

\renewcommand{\thepage}{A\arabic{page}}
\setcounter{page}{1}
\renewcommand{\thesection}{A\arabic{section}}
\setcounter{section}{0}
\renewcommand{\thetable}{A\arabic{table}}
\setcounter{table}{0}
\renewcommand{\theequation}{A\arabic{equation}}
\setcounter{equation}{0}

\begin{appendix}
\numberwithin{equation}{section}

\section{Stationary state of the isospectral model}
\label{app:steadystate}
Here we verify that the stationary state $\rho_{ss}$ in \cref{eq:stationary} is indeed stationary under evolution under $H_I$ and calculate its purity. For any time evolution operator with spectral decomposition $U=\sum_i e^{-it\lambda_i}\vert r_i\rangle\langle l_i\vert$ for $\lambda_i\in\mathbb{R}$, any state of the form $\rho = \sum_i p_i \vert r_i\rangle\langle r_i\vert$ is invariant in time. In the case of the isospectral TFIM in \cref{eq:isospectral-TFIM}, $\vert r_i\rangle=S(\beta)\vert i\rangle$, where $\vert i\rangle$ is an eigenvector of $H_\text{TFIM}$. Thus, $\rho_{ss}$ is one such state with $p_i=\text{Tr}[S(\beta)^2]^{-1}\ \forall i$. Analogous to the maximally mixed state $\mathbb{I}/d$, $\rho_{ss}$ is stationary regardless of choice of $H_\text{TFIM}$.

Using $U_I = S(\beta) (e^{-itH_\text{TFIM}}) S(\beta)^{-1}$, one can verify that $U_I S(\beta)^2 U_I^\dag = S(\beta)^2$, so
\begin{align}
    \rho_{ss}(t) = \frac{U_I \rho_{ss} U_I^\dag}{\operatorname{Tr}[U_I \rho_{ss} U_I^\dag]} = \frac{S(\beta)^2/\operatorname{Tr}[S(\beta)^2]}{\operatorname{Tr}\left[S(\beta)^2/ \operatorname{Tr}[S(\beta)^2]\right]} = \rho_{ss}
\end{align}
is stationary in time. The purity is computed using the fact that $\operatorname{Tr}[S(\beta)^{2a}] = \text{Tr}[\text{exp}(a\beta\sigma^z)]^L = (2\text{cosh}(a\beta))^L$,  to be
\begin{align}
    \label{eq:steadystate_purity}
    \operatorname{Tr}[\rho_{ss}^2] &= \frac{\left\Vert S(2\beta) \right\Vert_2^2}{\left\Vert S(\beta) \right\Vert_2^4 }= \frac{[2\text{cosh}(2\beta)]^L}{[2\text{cosh}(\beta)]^{2L}} = \left[\frac{1+\text{tanh}(\beta)^2}{2}\right]^L
\end{align}

These results generalize to any quasihermitian Hamiltonian, which can be written $H = S H_0 S^{-1}$ for some Hermitian $H_0$ and $S$. Furthermore, when restricted to the long-time eigenspace, \textit{any} non-Hermitian Hamiltonian will be quasihermitian up to an overall shift, and can be written in this form. Thus the forms \cref{eq:stationary} is valid for all non-Hermitian Hamiltonian evolutions in the long-time limit. 

\section{Alternative definition for a normalized OTOC}
\label{app:OTOC}

\subsection{Definition of the OTOC}
An alternative method of generalizing the OTOC to non-Hermitian systems relies on extending the notion of Heisenberg evolution. Motivated by the Heisenberg-Schrodinger correspondence of for unconditional (in general open-system) trajectories $\text{Tr}(A_t \rho) = \text{Tr}(A \rho_t)$, we propose defining a time evolved operator $A_t$ with respect to a state $\rho$ in terms of a CP time evolution superoperator $\mathcal{E}_t$ as
\begin{align}
    A_t \equiv \frac{\mathcal{E}_t^\dag(A)}{\text{Tr}(\mathcal{E}_t(\rho))}
\end{align}
where in the case that time evolution is generated by a non-Hermitian Hamiltonian, $\mathcal{E}_t(\cdot) = U_t \cdot U_t^\dag$ and $\mathcal{E}^\dag_t(\cdot) = U_t^\dag \cdot U_t$. Plugging this into the traditional OTOC (\cref{eq:OTOC-Hermitian}) for some yet arbitrary $\rho$ yields
\begin{align}
\begin{split}
\hat C_{V,W}(t,\rho) &\equiv
\frac{1}{d} \left[ \Vert V W_t \Vert_2^2 - \mathrm{Re}\text{Tr}( W^\dag_{t} V^\dag W_t V ) \right]\\
&=\label{eq:heis-OTOC}
\frac{1}{d\cdot\text{Tr}(\mathcal{E}_t(\rho))^2}\left[\left\Vert V \mathcal{E}_t^\dag(W) \right\Vert_2^2 - \mathrm{Re}\text{Tr}( \mathcal{E}_t^\dag(W^\dag) V^\dag \mathcal{E}_t^\dag(W) V )\right]
\end{split}
\end{align}

State dependence of the Heisenberg evolution arises from the nature of non-Hermitian evolutions as describing \textit{conditional} evolutions, which are conditional on the system state $\rho$. In the case where $\mathcal{E}_t$ is generated by a non-Hermitian Hamiltonian and $\rho = \mathbb{I}/d$ is the maximally mixed state, the denominator simplifies to 
\begin{align*}
    d \cdot\text{Tr}(U_t \rho U_t^\dag)^2 = d^{-1} \left\Vert U_t \right\Vert_2^4
\end{align*}

The numerator of \cref{eq:heis-OTOC} is the same as the unconditional open-system OTOC used in ~\cite{styliaris_information_2021, zanardi_information_2021-1}. Using similar techniques to those in the literature, we find the Haar averaged bipartite \textit{normalized} OTOC to be

\begin{align}
\begin{split}
    \hat G(t,\rho) &= \frac{d_B \text{Tr}(\text{Tr}_A (\mathcal{E}_t(\mathbb{I}))^2) - \text{Tr}( \mathbb{S}_{AA'} \mathcal{E}^{\otimes 2}_t(\mathbb{S}_{AA'}))}{d^2\cdot\text{Tr}(\mathcal{E}_t(\rho))^2 }\\
    &= \frac{d_B \text{Tr}(\text{Tr}_A (U_t U_t^\dag)^2) - \text{Tr}(U_t^{\dag \otimes 2} \mathbb{S}_{AA'} U_t^{\otimes 2} \mathbb{S}_{AA'})}{d^2\cdot\text{Tr}(U_t \rho U_t^\dag)^2 }
    \end{split}
\end{align}

Then in the case where $\mathcal{E}$ is generated by a non-Hermitian Hamiltonian and we take $\rho$ to be the maximally mixed state, we have

\begin{align}
\begin{split}
    \hat G(t,\mathbb{I}/d) &= \frac{d_B \text{Tr}(\text{Tr}_A (U_t U_t^\dag)^2)}{\left\Vert U_t \right\Vert_2^4}-\frac{\text{Tr}(U_t^{\dag \otimes 2} \mathbb{S}_{AA'} U_t^{\otimes 2} \mathbb{S}_{AA'})}{\left\Vert U_t \right\Vert_2^4}\\
    \label{eq:HABOTOC}
    &= E_{\mathrm{op}}^\text{lin}(U_t) - E_B(U_t U_t^\dag)
\end{split}
\end{align}

where 
\begin{align}
\label{eq:eB}
    E_B(U_t U_t^\dag) = 1 - d_B\frac{\text{Tr}(\text{Tr}_A (U_t U_t^\dag)^2)}{\left\Vert U_t \right\Vert_2^4}
\end{align} can be interpreted as a form of linear entanglement entropy on the $B$ subsystem, and reduces to 0 in the case $U_t$ is unitary. $E_{\mathrm{op}}^\text{lin}(U_t)$ is the linear operator entanglement of $U_t$ across the bipartition, as defined in Ref.~\cite{zanardi_entanglement_2001-1}, and reduces to the usual definition if $U_t$ is unitary.

\subsection{Long-time average}
\label{eq:OTOC-LTA}

The long-time average (LTA) of the bipartite Haar-averaged OTOC \cref{eq:HABOTOC} can also be taken via similar techniques to that of the 2-Renyi operator entanglement in Appendix~\ref{app:LTA-derivation}. For this we use the approximation $\overline{(f/g)} \approx \overline{f}/\overline{g}$
which is a weaker restriction than the one used in \cref{eq:LTA-Analytic}, and average $E_\mathrm{op}^\text{lin}$ and $E_B$ individually. 

$E_\mathrm{op}^\text{lin}(U_t)$ can be averaged identically to $E_\mathrm{op}(U_t)$ without the log, giving
\begin{align}
    \overline{E_\mathrm{op}^\text{lin}(U_t)} \approx 1- \frac{\operatorname{Tr}\left[ R_A L_A \right] + \operatorname{Tr}\left[ R_B L_B \right] - \operatorname{Tr}\left[ \mathrm{diag}(R_A) \mathrm{diag}(L_A) \right]}{[\text{Tr}[\eta_{_L}]]^2 + \text{Tr}[R L] - \text{Tr}[L]}.
\end{align}

The $E_B$ term (\cref{eq:eB}) has the same denominator as $E_\mathrm{op}^\text{lin}$, while the numerator may be expanded

\begin{align}
    \begin{split}
        \text{Tr}(\text{Tr}_A (U_t U_t^\dag)^2) &= \sum_{i,j,k,l} e^{it(\lambda_i^*+\lambda_j^*-\lambda_k-\lambda_l)}\operatorname{Tr}\left[\operatorname{Tr}_A(\vert r_k\rangle\langle l_k\vert l_i\rangle\langle r_i\vert )\operatorname{Tr}_A\left(\vert r_l\rangle\langle l_l\vert l_j\rangle\langle r_j\vert \right)\right]
    \end{split}
\end{align}
This term's LTA can be taken similarly to as in \cref{eq:LTA-numerator}, yielding

\begin{align}
    \overline{\text{Tr}(\text{Tr}_A (U_t U_t^\dag)^2)} &= (a) + (b) - (c)
\end{align}
for
\begin{align}
\begin{split}
    (a) &= \sum_{i,j\in\mathcal{H}_L} \operatorname{Tr}[\operatorname{Tr}_A(\vert r_i\rangle\langle r_i\vert)\operatorname{Tr}_A(\vert r_j\rangle\langle r_j\vert)]\langle l_i\vert l_i\rangle \langle l_j\vert l_j\rangle\\
    &= \sum_{i,j\in\mathcal{H}_L} \operatorname{Tr}[\rho_i^B \rho_j^B]\operatorname{Tr}[\sigma_i]\operatorname{Tr}[\sigma_j]\\
    &= \sum_{i,j\in\mathcal{H}_L}  (R_B)_{ij}(L_{AB})_{ij}\\
    &= \operatorname{Tr}[R_B L_{AB}]
    \end{split}
\end{align}
where $(L_{AB})_{ij} = \operatorname{Tr}[\sigma_i]\operatorname{Tr}[\sigma_j] = (\eta_{_L})_{ii}(\eta_{_L})_{jj}$ is the \textit{completely} reduced long-time left Gram matrix. For the second term we find,
\begin{align}\begin{split}
    (b) &= \sum_{i,j\in\mathcal{H}_L}  \operatorname{Tr}[\operatorname{Tr}_A(\vert r_i\rangle\langle r_j\vert)\operatorname{Tr}_A(\vert r_j\rangle\langle r_i\vert)]\langle l_i\vert l_j\rangle \langle l_j\vert l_i\rangle\\
    &= \sum_{i,j\in\mathcal{H}_L} \operatorname{Tr}[\mathbb{S}_{BB'}\vert r_i r_j \rangle \langle r_j r_i\vert ]\operatorname{Tr}[\sigma_i \sigma_j]\\
    &= \sum_{i,j\in\mathcal{H}_L} \langle r_i r_j\vert\mathbb{S}_{AA'}\vert r_i r_j \rangle L_{ij}\\
    &= \sum_{i,j\in\mathcal{H}_L}  (R_A)_{ij} L_{ij}\\
    &= \operatorname{Tr}[R_A L]
    \end{split}\end{align}
where the second line comes from $\text{Tr}[\text{Tr}_A(X)\text{Tr}_A(Y)] = \text{Tr}[\mathbb{S}\ \text{Tr}_A(X)\otimes\text{Tr}_A(Y)] =  \text{Tr}[\mathbb{S}_{BB'}(X\otimes Y)]$, the third from $\mathbb{S}_{AA'} = \mathbb{S}\mathbb{S}_{BB'}$, and the fourth line from \cref{eq:inner-to-Gram}. The final term is a special case of the second with $i=j$, yielding
\begin{align}
    \begin{split}
        (c) &= \sum_{i\in\mathcal{H}_L} (R_A)_{ii} L_{ii}\\
        &= \operatorname{Tr}[\text{diag}(R_A)\text{diag}(L)]
    \end{split}
\end{align}
Note that $\text{diag}(R_A)=\text{diag}(R_B)$ and $\text{diag}(L)=\text{diag}(L_{AB})$. This gives us the final form
\begin{align}
    \overline{E_B(U_t U_t^\dag)} \approx
    1-d_B \frac{\text{Tr}[R_B L_{AB}] + \text{Tr}[R_A L] - \text{Tr}[\mathrm{diag}(R_A)\mathrm{diag}(L)]}{[\text{Tr}[\eta_{_L}]]^2 + \text{Tr}[R L] - \text{Tr}[L]}
\end{align}

\section{Convergence of the OTOC}
\label{app:otoc-convergence}

\begin{figure}[!ht]
\includegraphics[width=0.9\linewidth]{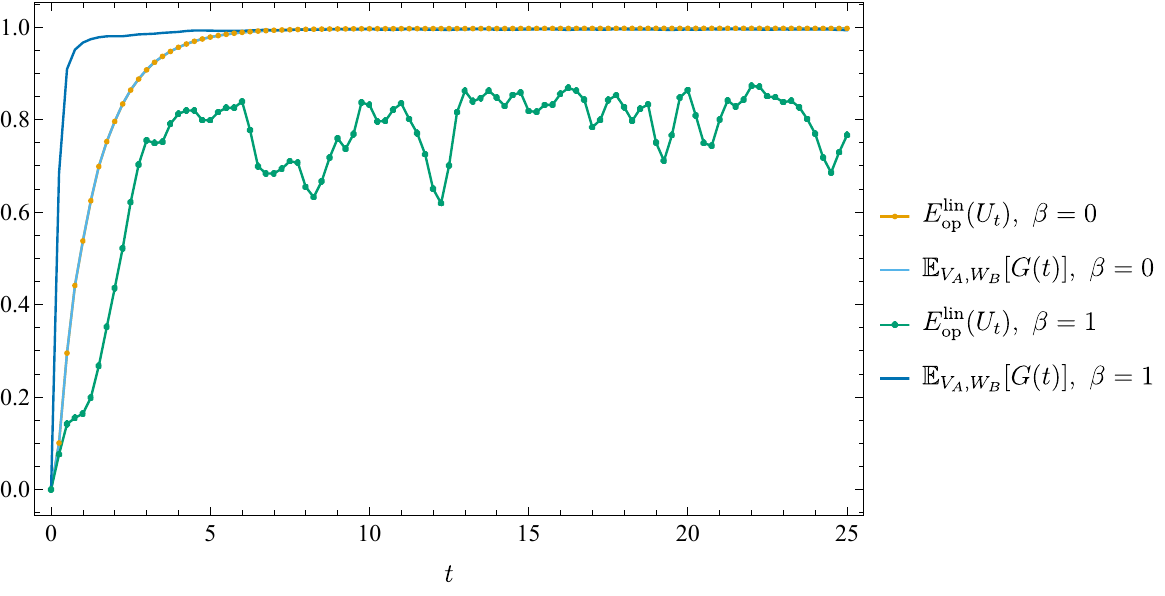}
\caption{Plots of the linear entanglement entropy (\cref{eq:linear-opent}) vs the Haar-averaged OTOC (\cref{eq:OTOC-namit}) for the isospectral chaotic TFIM at $L=10$. As expected, the operator entanglement and averaged OTOC are identical at $\beta=0$, the Hermitian case, but differ at $\beta=1$. The OTOC is averaged over $100$ pairs of unitaries $V_A,W_B$ drawn from the Haar distribution over their respective subsystem, at which point it has converged to low error. The isospectral model is used to demonstrate that the failure of the OTOC to converge to $E_\mathrm{op}^\text{lin}$ is due to the effect on eigenstates of $U_t$ rather than on the spectrum.}
\label{fig:OTOC-convergence}
\end{figure}

In the Hermitian case, the OTOC in \cref{eq:OTOC-namit} averaged over sets of Haar-random unitaries $V_A, W_B$ converges to linear entanglement entropy of $U_t$, given by
\begin{align}
    E_\mathrm{op}^\text{lin} = 1-\frac{1}{\Vert U_t \Vert_2^4}\text{Tr}\left[ \mathbb{S}_{AA'} U_t^{\otimes 2} \mathbb{S}_{AA'} U_t^{\dag \otimes 2} \right].
    \label{eq:linear-opent}
\end{align}
As shown in \cref{fig:OTOC-convergence}, this no longer holds for non-Hermitian systems, for which the OTOC saturates more quickly even as operator entanglement drops. 

\section{Operator entanglement from the Choi state}
\label{app:choi}
We show here that the form of operator entanglement used in \cref{eq:2-renyi-ent} is motivated directly by normalization of the Choi state. 

For CP channel $\mathcal{E}$ the associated Choi state is 
\begin{align}
    C(\mathcal{E}) = \frac{1}{d}\sum_{i,j=1}^d (\mathcal{E}\otimes\mathbb{I})(\vert ii\rangle\langle jj\vert).
\end{align}
which can be interpreted as a normalized state either if $\mathcal{E}$ is trace-preserving or if we explicitly divide by the trace $\text{Tr}[C(\mathcal{E})] = \text{Tr}[\mathcal{E}(\mathbb{I})]$.

The operator entanglement $E_\text{op}(\mathcal{E})$ across a bipartition can be defined as the entanglement entropy (across the same bipartition) of the Choi state \cite{Li_2021}. For e.g. the linear entropy across an A$\vert$B bipartition, that is, \begin{align}
E^\text{lin}_\text{op}(\mathcal{E}) &= 1-\frac{\text{Tr}[\text{Tr}_A[C(\mathcal{E})]^2]}{\text{Tr}[C(\mathcal{E})]^2}\\
&= 1-\frac{\text{Tr}[S_{AA'}\mathcal{E}^{\otimes 2}(S_{AA'})]}{ \text{Tr}[\mathcal{E}(\mathbb{I})]^2}
\end{align}
where the numerator comes from the Choi-Jamiołkowski isomorphism. For the pertinent case $\mathcal{E}(\cdot)=U\cdot\ U^\dag$, this becomes the standard form
 \begin{align}
     E^\text{lin}_\text{op}(\mathcal{E}) = 1-\frac{\text{Tr}[S_{AA'}U^{\otimes 2}S_{AA'}U^{\dag \otimes 2}]}{\Vert U\Vert_2^4}
 \end{align}

\section{Analytic form of $\overline{E_\mathrm{op}(U_t)}$}
\subsection{Derivation of the analytical formula}
\label{app:LTA-derivation}

We wish to prove \cref{eq:LTA-Analytic}, that is, to find an analytic approximation to the long time average of the equation

\begin{align}
    \label{eq:renyi-2}
    E_\mathrm{op}(U_t) = -\text{log}\left( \frac{\operatorname{Tr}\left[ \mathbb{S}_{AA'} U_t^{\otimes 2} \mathbb{S}_{AA'} U_t^{\dagger \otimes 2} \right]}{\left\Vert U_t \right\Vert_{2}^{4}} \right).
\end{align}

The approximation $\overline{\text{log}(f)} \approx \text{log}(\hspace{1mm}\overline{f }\hspace{1mm})$ allows the time average to be applied to the numerator and denominator of \cref{eq:renyi-2} separately and within the logarithm, using $\text{log}(f/g)=\text{log}(f)-\text{log}(g)$. Using the spectral decomposition $U = \sum_i e^{-it\lambda_i}\vert r_i\rangle\langle l_i\vert$, the numerator within the logarithm may be expanded as 
\begin{align}
\operatorname{Tr}\left[ \mathbb{S}_{AA'} U^{\dagger \otimes 2}_{t} \mathbb{S}_{AA'} U^{\otimes 2}_{t}\right] &= \sum_{i,j,k,l} e^{it(\lambda_i^*+\lambda_j^*-\lambda_k-\lambda_l)} \operatorname{Tr}\left( \mathbb{S}_{AA'} \vert l_i l_j\rangle\langle r_i r_j\vert\mathbb{S}_{AA'}\vert r_k r_l\rangle\langle l_k l_l\vert \right)
 \end{align}.
 
 As discussed in the main text, we can assume without loss of generality that $\text{max}\{\text{Im}[\lambda_i]\}=0$, and when taking the LTA drop all terms in the sum with $\text{Im}[\lambda_i]<0$, as in the limit $t\rightarrow\infty$ they will be zero, leaving only terms in $\mathcal{H}_L$. The numerator LTA is then
 
 \begin{align}
 \label{eq:LTA-numerator}
 \begin{split}
     \overline{ \operatorname{Tr}\left[ \mathbb{S}_{AA'} U^{\dagger \otimes 2}_{t} \mathbb{S}_{AA'} U^{\otimes 2}_{t}\right] } &= \sum_{i,j,k,l\in\mathcal{H}_L} \overline{e^{it(\lambda_i+\lambda_j-\lambda_k-\lambda_l)}} \langle r_i r_j\vert\mathbb{S}_{AA'}\vert r_k r_l\rangle \langle l_k l_l \vert \mathbb{S}_{AA'} \vert l_i l_j\rangle\\
     &= \sum_{i,j,k,l\in\mathcal{H}_L} (\delta_{ik}\delta_{jl}+\delta_{il}\delta_{jk}-\delta_{ijkl})\langle r_i r_j\vert\mathbb{S}_{AA'}\vert r_k r_l\rangle \langle l_k l_l \vert \mathbb{S}_{AA'} \vert l_i l_j\rangle\\
     &= (a)+(b)-(c)
     \end{split}
 \end{align}

where in the third line we leverage use of the NRC to take the time average, as in Ref.~\cite{styliaris_information_2021}, and $\delta$ are Kronecker deltas. The three terms are
\begin{align}
(a) = \sum\limits_{i,j\in\mathcal{H}_L}  \left\langle r_i r_j | \mathbb{S}_{AA'} | r_i r_j \right\rangle \left\langle l_i l_j | \mathbb{S}_{AA'} | l_i l_j \right\rangle
\end{align}
\begin{align}
(b) = \sum\limits_{i,j\in\mathcal{H}_L}  \left\langle r_i r_j | \mathbb{S}_{BB'} | r_i r_j \right\rangle \left\langle l_i l_j | \mathbb{S}_{BB'} | l_i l_j \right\rangle
\end{align}
\begin{align}
(c) = \sum\limits_{i\in\mathcal{H}_L}  \left\langle r_i r_i | \mathbb{S}_{AA'} | r_i r_i \right\rangle \left\langle l_i l_i | \mathbb{S}_{AA'} | l_i l_i \right\rangle
\end{align}

Note that the final term is simply the product of the purities of the left/right eigenvectors, and is symmetric under the swap $A\leftrightarrow B$. Each term can be written in terms of modified Gram matrices using the definitions in \cref{eq:reduced-Gram} and \cref{eq:reduced-eigenstates} and techniques from Ref.~\cite{styliaris_information_2021}:
\begin{align}
\label{eq:inner-to-Gram}
\begin{split}
    \left\langle l_i l_j | \mathbb{S}_{AA'} | l_i l_j \right\rangle &= \text{Tr}\left[\text{Tr}_{B}\left(\vert l_i\rangle\langle l_i\vert\right)\text{Tr}_{B}\left(\vert l_j\rangle\langle l_j\vert\right)\right]\\
    &= \langle \sigma^A_i, \sigma^A_j\rangle_A\\
    &\equiv (L_A)_{ij}
    \end{split}
\end{align}
and similarly under the swap $\{\vert l\rangle,\sigma^A,L_A\} \leftrightarrow \{\vert r\rangle,\rho^A,R_A$\}, and the swap $A \leftrightarrow B$. This gives, for example, 
\begin{align*}
    (a) = \sum_{ij\in\mathcal{H}_L} (R_A)_{ij} (L_A)_{ij}= \text{Tr}[R_A L_A]
\end{align*}

Using NRC, the time average of the denominator is then
\begin{align}
\begin{split}
\overline{ \left\Vert U_{t} \right\Vert_{2}^{4} } &= \overline{\operatorname{Tr}\left[ U^{\dagger \otimes 2}_{t} U^{\otimes 2}_{t} \right]}\\
&= \sum_{i,j,k,l\in\mathcal{H}_L} \overline{e^{it(\lambda_i+\lambda_j-\lambda_k-\lambda_l)}} \langle r_i r_j\vert r_k r_l\rangle\langle l_k l_l \vert l_i l_j\rangle\\\\
&= (d) + (e) - (f).
\end{split}
\end{align}
Here,
\begin{align}
\begin{split}
(d) &= \sum\limits_{i,j\in\mathcal{H}_L} \langle r_i\vert r_i\rangle\langle r_j\vert r_j\rangle\langle l_i\vert l_i\rangle\langle l_j\vert l_j\rangle\\
&= (\sum\limits_{i\in\mathcal{H}_L} \langle l_i\vert l_i\rangle\ )^2\\
&= [\text{Tr}[\eta_{_L}]]^2
\end{split}
\end{align}
where we use the normalization $\langle r_i\vert r_i\rangle = 1\ \forall i$. Similarly,
\begin{align}
\begin{split}
(e) &= \sum\limits_{i,j\in\mathcal{H}_L} \left| \left\langle r_i | r_j \right\rangle \right|^{2} \left| \left\langle l_i | l_j \right\rangle \right|^{2}\\
&= \sum\limits_{i,j\in\mathcal{H}_L}\text{Tr}\left[\rho_i \rho_j\right]\text{Tr}\left[\sigma_i \sigma_j\right]\\
&= \sum\limits_{i,j\in\mathcal{H}_L} R_{ij}L_{ij}\\
&= \text{Tr}\left[R L\right]
\end{split}
\end{align}
where $\rho$, $\sigma$, $L$, and $R$ are the unreduced long-time eigenstates and Gram matrices. Note $L_{ij}=\vert(\eta_{_L})_{ij}\vert^2$. Finally,
\begin{align}
\begin{split}
(f) &= \sum\limits_{i\in\mathcal{H}_L} \vert \langle l_i\vert l_i \rangle \vert^{2}\\
&= \sum\limits_{i\in\mathcal{H}_L} L_{ii}\\
&= \text{Tr}\left[L\right]
\end{split}
\end{align}
where we have again dropped $\left\Vert r_i\right\Vert = 1$. Notice that all three terms manifest the non-Hermiticity of the Hamiltonian in the eigenvector non-orthonormality.

\subsection{Verifying accuracy of analytic form}
\label{app:LTA-analytic-plots}

\begin{figure}
\raggedright
    \includegraphics[width=1\linewidth]{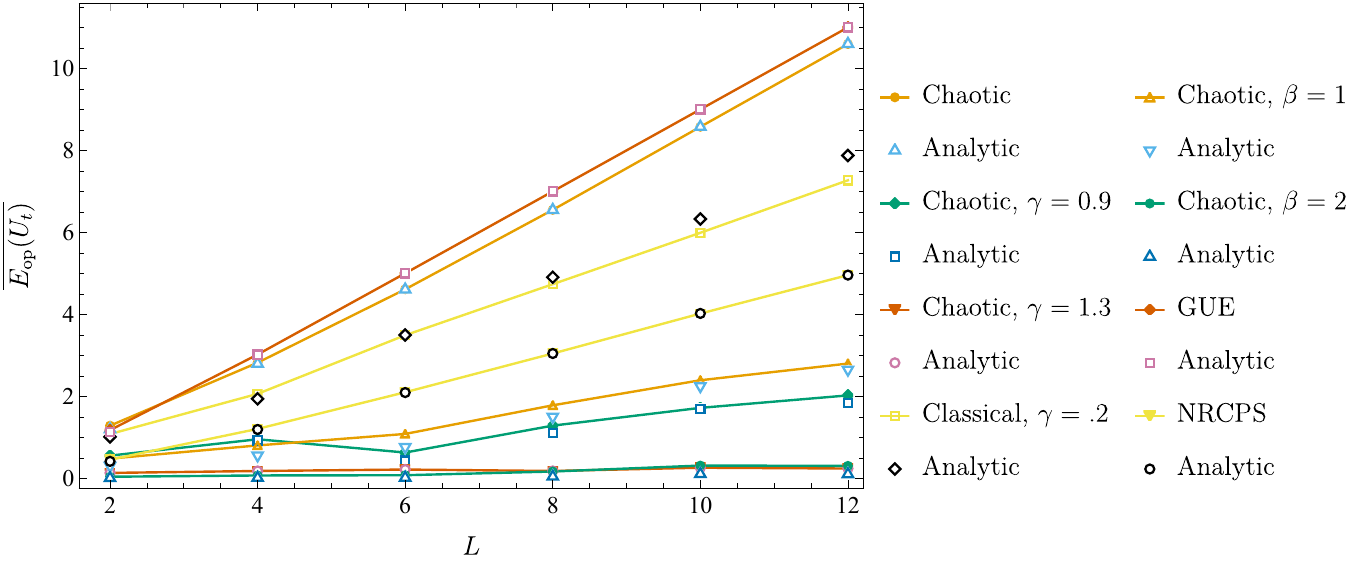}
    \caption{Comparison of numerically and analytically calculated values of the operator entanglement LTA as a function of system size $L$ for a range of models. The numeric values are the same as in \cref{fig:scaling_comparison}, but the integrable and Hermitian classical models are omitted for visibility as they do not satisfy NRC. The analytic approximation has visible error for the classical $\gamma=0.2$ and chaotic $\beta=1$ models, but in this case still captures the correct scaling behavior.}
    \label{fig:scaling-comp-analytic}
\end{figure}

\begin{figure}
\raggedright
    \includegraphics[width=1\linewidth]{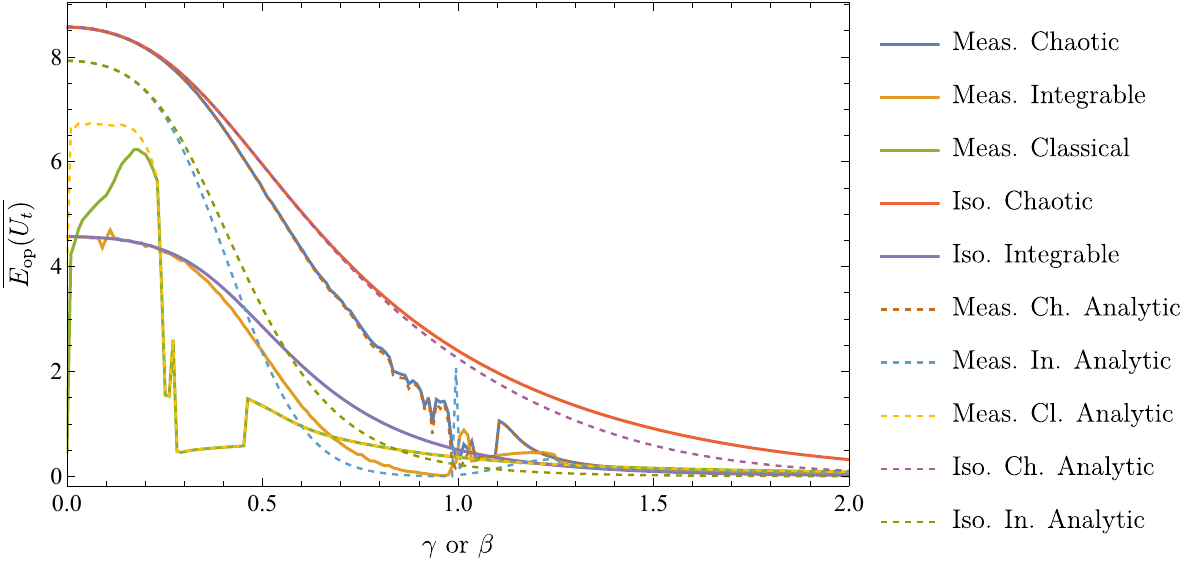}
    \caption{Comparison of numerically and analytically calculated operator entanglement LTA as a function of nonhermiticity parameter for a range of models. Numerical values are the same as in \cref{fig:param_scaling}. Note large deviations come from the integrable TFIM and near-Hermitian classical TFIM, which violate NRC (or in the classical TFIM with $0<\gamma<.2$, nearly violate) but are included for completeness.}
    \label{fig:param-scaling-analytic}
\end{figure}

By comparison of numerically and analytically calculated operator entanglement averages in \cref{fig:scaling-comp-analytic}, we see that for the NRC-satisfying models and nonhermiticity parameters considered it is highly accurate in most cases, and sufficiently accurate to capture the correct scaling behavior in the rest. From the comparison in \cref{fig:param-scaling-analytic}, we see that the analytic approximation is primarily inaccurate in the cases where NRC is not satisfied: for the classical and chaotic models at their degeneracy points ($\gamma=1$ and $\gamma=0$ respectively) and for the integrable models in general. However away from points where NRC is not satisfied, the analytic approximation does capture all interesting behavior.

\end{appendix}

\bibliographystyle{apsrev4-1}
\bibliography{library,refs}

\end{document}